\documentclass[iop]{emulateapj}

\usepackage{amsmath}

\newcommand{\jband}{$J_{125}$}
\newcommand{\hband}{$H_{160}$}
\newcommand{\yband}{$Y_{098}$}
\newcommand{\vband}{$V$}

\usepackage{hyperref}

\shorttitle{$z \sim 10$ galaxy candidates in BoRG[\lowercase{z}8]}
\shortauthors{Bernard et al.}

\begin{document}

\title{Galaxy candidates at $\lowercase{z} \sim 10$ in archival data from the \\ 
Brightest of Reionizing Galaxies (BoRG[\lowercase{z}8]) survey}

\author{S. R. Bernard\altaffilmark{1}\altaffilmark{,2}, D. Carrasco\altaffilmark{1}, M. Trenti\altaffilmark{1}, P. A. Oesch\altaffilmark{3}\altaffilmark{,4}, J. F. Wu\altaffilmark{5}, L. D. Bradley\altaffilmark{6}, K. B. Schmidt\altaffilmark{7}\altaffilmark{,8}, R. J. Bouwens\altaffilmark{9}, V. Calvi\altaffilmark{6}, C. A. Mason\altaffilmark{7}\altaffilmark{,10},  M. Stiavelli\altaffilmark{6}, T. Treu\altaffilmark{10}}

\altaffiltext{1}{School of Physics, The University of Melbourne, Parkville, VIC 3010, Australia. Email: \url{sbernard@student.unimelb.edu.au}}
\altaffiltext{2}{ARC Centre of Excellence for All-sky Astrophysics (CAASTRO).}
\altaffiltext{3}{Yale Center for Astronomy and Astrophysics, Physics Department, New Haven, CT 06520, USA.}
\altaffiltext{4}{Department of Astronomy, Yale University, New Haven, CT 06520, USA.}
\altaffiltext{5}{Department of Physics and Astronomy, Rutgers, The State University of New Jersey, 136 Frelinghuysen Road, Piscataway, NJ 08854, USA}
\altaffiltext{6}{Space Telescope Science Institute, 3700 San Martin Drive, Baltimore, MD 21218, USA.}
\altaffiltext{7}{Department of Physics, University of California, Santa Barbara, CA 93106-9530, USA.}
\altaffiltext{8}{Leibniz-Institut fur Astrophysik Potsdam (AIP), An der Sternwarte 16, 14482 Potsdam, Germany.}
\altaffiltext{9}{Leiden Observatory, Leiden University, NL-2300 RA Leiden, The Netherlands.}
\altaffiltext{10}{Department of Physics and Astronomy, UCLA, Los Angeles, CA, 90095-1547, USA.}

\begin{abstract}

  The Wide Field Camera 3 (WFC3) on the \emph{Hubble Space Telescope}
  (\emph{HST}) enabled the search for the first galaxies observed at
  $z\sim 8-11$ ($500- 700$ Myr after the Big Bang). To continue
  quantifying the number density of the most luminous galaxies
  ($M_{AB}\sim -22.0$) at the earliest epoch observable with \emph{HST}, we
  search for $z\sim 10$ galaxies (F125W-dropouts) in archival data
  from the Brightest of Reionizing Galaxies (BoRG[z8]) survey,
  originally designed for detection of $z\sim 8$
  galaxies (F098M-dropouts). By focusing on the deepest 293 arcmin$^2$ of the data
  along 62 independent lines of sight, we identify six $z\sim 10$
  candidates satisfying the color selection criteria, detected at S/N
  $>8$ in F160W with $M_{\mathrm{AB}} = -22.8$ to $-21.1$ if at
  $z = 10$. Three of the six sources, including the two brightest, are
  in a single WFC3 pointing ($\sim 4$ arcmin$^2$), suggestive of
  significant clustering, which is expected from bright galaxies at
  $z\sim10$. However, the two brightest galaxies are too extended to
  be likely at $z\sim10$, and one additional source is unresolved and
  possibly a brown dwarf.  The remaining three candidates have
  $m_{AB}\sim 26$, and given the area and completeness of our search,
  our best estimate is a number density of sources that is marginally
  higher but consistent at $2\sigma$ with searches in legacy
  fields. Our study highlights
  that $z\sim10$ searches can yield a small number of candidates, making tailored follow-ups of
  \emph{HST} pure-parallel observations viable and effective.

\end{abstract}



\section{Introduction}

The epoch of reionization signified the appearance of the first
stars and galaxies within the first billion years after the Big Bang, and the transformation of the intergalactic medium (IGM) from
opaque to transparent. Despite recent
progress, however, it is not yet fully understood. It is now well established that reionization is
completed by $z\sim 6$ thanks to observations of the Ly$\alpha$ forest (e.g. 
\citealt{willott07}), and that the Universe was substantially
ionized around redshift $z\sim 8$ when its age was less than 600 Myr,
based on the electron scattering optical depth measured by Planck
\citep{planck15}. However, there is still
substantial uncertainty regarding the sources of reionization. Can
galaxies form with sufficient efficiency at such early times to
provide enough reionizing photons (e.g. \citealt{alvarez12}), or is
the process possibly driven by other classes of objects such as AGN
\citep{giallongo12, madau15}?

Observationally, recent progress in near-IR detector technology has
dramatically advanced our ability to search for galaxies during this
epoch. Following the installation of the Wide Field Camera 3 (WFC3) on
the \emph{Hubble Space Telescope} (\emph{HST}), a continuously growing
sample of galaxy candidates at $z\gtrsim 7$ is accumulating thanks to
a variety of surveys. These range from small-area ultradeep observations such as the Hubble Ultra-deep Field (HUDF,
\citealt{illingworth13}), to shallower, larger-area searches for
$L\gtrsim L_*$ galaxies either in legacy fields such as the Cosmic Assembly Near-infrared Deep Extragalactic Legacy Survey (CANDELS;
\citealt{grogin11, koekemoer11}), or taking advantage of random-pointing
opportunities like in the Brightest of Reionizing Galaxies
(BoRG) survey (GO 11700, 12572, 13767; PI Trenti). Overall, a sample
approaching 1000 galaxy candidates at $z>7$ is known today
\citep{bouwens15a}, and we are beginning to identify
the first galaxy candidates from the first 500 million years
($z \sim 9-10$; \citealt{bouwens11, bouwens14,bouwens15a, bouwens15b, zheng12, coe13, ellis13, mclure13, oesch14,zheng14,zitrin14,mcleod15,ishigaki15,infante15}). 

These observations provide solid constraints on the galaxy luminosity
function (LF) out to $z\sim 8$, which appears to be overall well
described by a~\citet{schechter76} form,
$\Phi(L) = \Phi^* (L/L^*)^\alpha \exp{(-L/L^*)}/L^*$, as at lower
redshift \citep{bouwens07,schmidt14,bouwens15a,finkelstein15}. However, other studies suggest
that bright galaxy formation might not be suppressed as strongly at
$z\gtrsim 7$, and either a single power law
\citep{bouwens11b,finkelstein15} or a double power law
\citep{bowler14} fit to the bright end of the LF has been
explored. This change in the shape of the bright end is in turn
connected theoretically to the physics of star formation in the most
overdense and early forming environments where the brightest and
rarest galaxies are expected to live \citep{munoz_loeb08,trenti12}. A
departure from a Schechter form could indicate a lower efficiency of
feedback processes at early times, which in turn would imply an
increase in the production of ionizing photons by
galaxies. Additionally, at $z \ge 8$, the observed number density of
bright galaxies is affected by magnification bias
\citep{wyithe11,baronenugent15a,mason15,fialkov15}, and this bias can
cause the LF to take on a power-law shape at the bright
end. Currently, the samples at $z\gtrsim 9$ are still too small to
draw any conclusion on which scenario is realized, since only a
handful of $z\sim 9-10$ candidates are known.

In addition to constraining the shape of the LF, the
brightest high-$z$ candidates identified by \emph{HST} observations are also
ideal targets for follow-up observations to infer stellar population
properties such as ages and stellar masses \citep{stark09,labbe10,labbe15,grazian15},
ionization state of the IGM \citep{munoz11}, and spectroscopic
redshift. For the latter, confirmation of
photometric candidates relies typically on detection of a Lyman break
in the galaxy continuum, (e.g., \citealt{malhotra05}) and/or of emission lines,
primarily Lyman-$\alpha$ (e.g. \citealt{stark10,pentericci11,pentericci14,caruana12,schenker12,treu12,treu13,finkelstein13,tilvi14,vanzella14}) or other UV lines such as
CIII] or CIV \citep{stark15a, stark15b}. Spectroscopic follow-up for
sources at $z\gtrsim 7.5$ is extremely challenging, with only limits
on line emission resulting from most observations. Yet, the brightest targets show significant
promise of detection based on the latest series of follow-ups which
led to spectroscopic confirmation out to $z=8.7$ \citep{zitrin15}, with several other Ly$\alpha$ detections at
$z\gtrsim 7.5$ \citep{oesch15,robertsborsani15}.

With the goal of complementing the discovery of the rarest and most
luminous sources in the epoch of reionization from legacy fields such
as CANDELS, the Brightest of Reionizing Galaxies Survey (BoRG, see
\citealt{trenti11}) has been carrying out pure-parallel, random
pointing observations with WFC3 since 2010. BoRG identified a large
sample ($n=38$) of $z\sim 8$ $Y$-band dropouts with $L\gtrsim L_*$
(\citealt{trenti11, trenti12,bradley12,schmidt14}; see also
\citealt{mclure13,bouwens15a}). This represents a catalog of galaxies
that is not affected by large scale structure bias (sample or
``cosmic'' variance; see \citealt{trenti08}), which is especially severe
for rare sources sitting in massive dark matter halos
($M_{h}\gtrsim 10^{11}~\mathrm{M_{\odot}}$), as inferred from
clustering measurements at $z>7$ \citep{baronenugent14}. Follow-up
spectroscopy of the BoRG dropouts with Keck and VLT has provided
evidence for an increase of the IGM neutrality at $z\sim 8$ compared
to $z\sim 6-7$ \citep{treu12,treu13,baronenugent15}. Currently, a new
campaign of observations is ongoing, with a revised filter-set
optimized for the new frontier of redshift detection at $z\sim 9-10$
(BoRG[z9-10]; GO 13767, PI Trenti). Initial results from $\sim 25\%$ of the dataset
($\sim 130$ arcmin$^2$) led to the identification of two candidates at
$z\sim 10$ \citep{calvi15} with $m_{\mathrm{AB}}\sim 25-25.5$, which
are similar in luminosity to the spectroscopically confirmed $z=8.7$
source reported by \citet{zitrin15}, but significantly brighter than
the six $J$-dropouts with $m_{\mathrm{AB}}\sim 26-27$ identified in the
GOODS/CANDELS fields from a comparable area \citep{oesch14}.

These recent developments indicate that it might be possible for a
small number of ultra-bright sources ($M_{\mathrm{AB}} \lesssim -22$)
to be present as early as 500 Myr after the Big Bang. Thus, they
prompted us to systematically analyze the BoRG archival data from
observations in the previous cycles, which cover $\sim 350$
arcmin$^2$, to search for bright $z\sim 10$ candidates and constrain
their number density. This paper presents the results of this search, and is organized as follows: Section
\ref{section2} briefly introduces the BoRG dataset. Section
\ref{section3} discusses our selection criteria for $z\sim 10$ sources
(\jband-band dropouts), with results presented in
Section~\ref{section4}. In Section \ref{lfsection}, we determine the
galaxy UV luminosity function at $z \sim 10$, and compare with
previous determinations. Section~\ref{sec:conclusions} summarizes and
concludes.  Throughout the paper we use the \citet{planck15}
cosmology: $\Omega_{\Lambda} = 0.692$, $\Omega_M = 0.308$ and
$H_0 = 67.8\,\mathrm{km}\,\mathrm{s}^{-1}\,\mathrm{Mpc}^{-1}$. All
magnitudes are quoted in the AB system \citep{oke83}.

\section{The BoRG survey}\label{section2}

We use data acquired as part of the Brightest of Reionizing Galaxies
(BoRG[z8]) survey, which consists of core BoRG pointings (GO 11700, 12572,
12905), augmented by other pure parallel archival data (GO
11702, PI Yan, \citealt{yan11}) and COS GTO coordinated parallel observations. For an
in-depth description of the survey, we refer the reader to
\citet{trenti11, bradley12, schmidt14}. Here, we use the 2014 (DR3)
public release of the
data\footnote{https://archive.stsci.edu/prepds/borg/}, which
consists of 71 independent pointings
covering a total area of $\sim$350 arcmin$^2$. All fields were imaged
using the WFC3/IR filters F098M, F125W and F160W, and in the optical
$V$ band, using either the WFC3 F606W or F600LP filter. We refer to
the WFC3 F098M, F125W and F160W images as the \yband, \jband{} and
\hband{} images, and to the F606W and F600LP images as $V_{606}$ and $V_{600}$, respectively.

Exposure times in each filter vary on a field-by-field basis, and
5$\sigma$ limiting magnitudes for point sources and aperture $r=0\farcs2$
are between $m_{\mathrm{AB}} = 25.6 - 27.4$, with a typical value of
$m_{\mathrm{AB}} \sim 26.7$ \citep{trenti11, bradley12, schmidt14}. We
note that since the dataset originates from parallel observations when
the primary instrument is a spectrograph (COS or STIS), there is no
dithering of the exposures. To compensate for the lack of dithering,
the BoRG data reduction pipeline has been augmented with a customized
Laplacian edge filtering algorithm developed by
\citet{vandokkum01}. Overall, the lack of dithering has a minimal
impact ($\Delta m< 0.1$) on the image and photometric quality, as it
has been established through comparison between primary (dithered)
versus pure-parallel observations of the same field \citep{calvi15}.

\begin{figure}
\epsscale{1.15}
\plotone{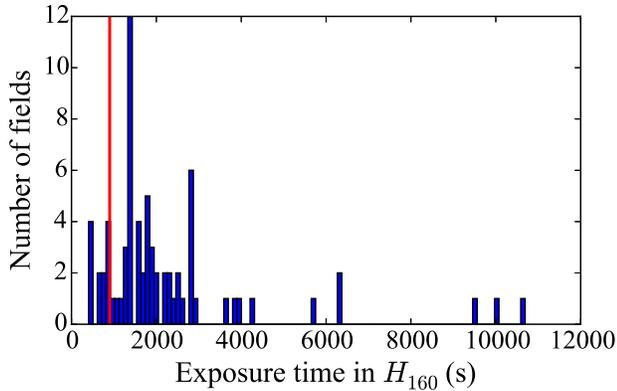}
\caption{Histogram of the exposure time in $H_{160}$ for the 71
  BoRG[z8] fields. The vertical red line indicates $t_{\mathrm{exp}} =
  900$ s. Fields with exposure times $< 900$ s in $H_{160}$ were
  excluded from our analysis.\label{fieldhist}}
\end{figure}

Since the BoRG[z8] survey was designed to have \jband{} as primary
detection band, some fields have only a single short exposure in the
\hband-band. To ensure a consistently high image quality, here we
include in the analysis only those fields with total exposure time
$t_\mathrm{exp}\geq 900~\mathrm{s}$ in \hband. This resulted in the exclusion
of 9 fields out of 71, so that the area included in our study
is $293$ arcmin$^2$. The distribution of the exposure time in \hband{} for the fields in BoRG[z8] is shown in Figure~\ref{fieldhist}.

The BoRG[z8] public data release consists of reduced and aligned
science images produced with \texttt{MultiDrizzle} \citep{koekemoer03} with a
pixel scale of $0\farcs08$, as well as associated weight maps (see
\citealt{bradley12,schmidt14}). Following our standard analysis
pipeline to search for dropouts in the data \citep{trenti11}, we
create RMS maps from the weight maps, and normalize them to account
for correlated noise induced by \texttt{MultiDrizzle} (see
\citealt{casertano00}). In short, this is done for
each field and filter by measuring the noise in the image at random
positions not associated with detected sources (i.e. the ``sky''
noise), and comparing the measurement with the value inferred from the RMS map, which
can then be corrected by a multiplicative factor to match the
measurement. Our rescaling factors are on average $\sim 1.1$ for
the IR filters and $\sim 1.4$ for \vband{} (see  also \citealt{trenti11,bradley12,schmidt14}). In addition, photometric
zero-points are corrected to account for galactic reddening along each
line of sight, according to \citet{schlafly11}.

Using \texttt{SExtractor} \citep{bertin96} in dual-image mode, we construct catalogues of
sources in each field, using the \hband-band image for
detection. Colors and signal to noise ratios are defined based on
isophotal fluxes/magnitudes (FLUX\_ISO), while we adopt MAG\_AUTO for
the total magnitude of each source.


\section{Selection of $J_{125}$-dropouts}\label{section3}

To select $z \sim 9-10$ galaxy candidates, we use the dropout
technique \citep{steidel96}. At high $z$, neutral hydrogen in the
IGM almost completely absorbs UV photons, leading to
a break at the galaxy rest wavelength of Ly$\alpha$ (1216
\AA). For galaxies between $z \sim 9-11$, this implies a drop in
the \jband-filter, and non-detection in the \vband{} and \yband{} bands. 

Our focus on \jband-dropouts implies that our sample of candidates are
essentially detected only in \hband. Therefore, to minimize the risk
of introducing spurious sources, we require a clear detection in
\hband, with S/N$_{H} \geq 8$. We also impose a strong
$J_{125}-H_{160} $ break, trading sample completeness for higher
purity, and require a color cut: $J_{125}-H_{160} > 1.5$, which is
more conservative than the typical $J_{125}-H_{160} > 1.2$ applied to
legacy fields (e.g. \citealt{bouwens15a}) since we do not have the
availability of multi-observatory data to constrain the continuum of
candidates at longer wavelengths and help control contamination.

Overall, we impose the following criteria for selection as
\jband-dropouts:
\begin{multline*}\\
	J_{125} - H_{160} > 1.5 \\
	\mathrm{S}/\mathrm{N}_{V} < 1.5 \\
	\mathrm{S}/\mathrm{N}_{098} < 1.5\\
	\mathrm{S}/\mathrm{N}_{160} \geq 8.0 \\
\end{multline*}

When computing the $J_{125}-H_{160}$ color, if the \jband-band flux has
  S/N $<1$ we use the $1\sigma$ limit instead. 
  
  Finally, to reduce the risk of contamination from detector defects
surviving the data reduction pipeline, we further impose a stellarity
cut through the \texttt{SExtractor} CLASS\_STAR parameter. We require
CLASS\_STAR $<0.95$, where 1 corresponds to a point source, and 0 to a
diffuse light profile. We then visually inspected the dropouts that meet these criteria to reject any remaining detector artifacts and diffraction spikes. All the sources that meet all criteria and pass the visual inspection are
listed in Table \ref{table1}, and discussed below.

\section{Results}\label{section4}

\begin{deluxetable*}{lccccccccccc}
\centering
\tabletypesize{\footnotesize}
\tablecolumns{12}
\tablewidth{0pt}
\tablecaption{Photometry of \jband-dropouts \label{table1}}
\tablehead{ \colhead{ID}  & \colhead{RA (deg)}  &
  \colhead{Decl. (deg)} &  \colhead{\hband}  & \colhead{\jband $-$ \hband}  & \colhead{S/N$_{H}$}  & \colhead{S/N$_{J}$} & \colhead{S/N$_{Y}$}  & \colhead{S/N$_{V}$} & \colhead{$r_e$} & \colhead{Stellarity} & $M_{\mathrm{AB}}$\tablenotemark{a}}
\startdata
borg\_0240-1857\_25& 40.1195 & -18.9726 & 26.24 $\pm$ 0.18 & $ > 2.53$ & 8.1 & $ 0.1 $ & -0.1 & 1.0\tablenotemark{b} & $0\farcs13$ & 0.71 & $-21.1$\\
borg\_0240-1857\_129 & 40.1289 & -18.9678 & 24.74 $\pm$ 0.07 & $ 2.21$ & 14.5 & 2.5 & 0.6 & 0.9\tablenotemark{b} & $0\farcs33$ & 0.02 & $-22.7$\\
borg\_0240-1857\_369  & 40.1274 & -18.9612 & 25.22 $\pm$ 0.11 & $ 1.88$ & 9.6 & 2.2 & -1.7 & 0.2\tablenotemark{b} & $0\farcs38$ & 0.00 & $-22.3$\\
borg\_0456-2203\_1091  & 73.9774 & -22.0372 & 26.09 $\pm$ 0.13 & $> 2.47$ & 8.1 & -1.3 & -0.4 & 0.1\tablenotemark{c}& $0\farcs24$ & 0.51 & $-21.4$\\
borg\_1153+0056\_514 & 178.1972 & 0.9270 & 26.31 $\pm$ 0.24 & $>2.64$ & 8.0 & 0.02 & -0.1 & -0.6\tablenotemark{c} & $0\farcs23$ & 0.01 & $-21.2$\\
borg\_1459+7146\_785 & 224.7239 & 71.7814 & 25.82 $\pm$ 0.14 & 1.57 & 12.8 & 3.7 & -1.1 & 1.3\tablenotemark{c} & $0\farcs14$ & 0.91 & $-21.5$
\enddata
\tablenotetext{a}{Assuming $z = 10$.}
\tablenotetext{b}{$V_{600}$}
\tablenotetext{c}{$V_{606}$}
\end{deluxetable*}
\begin{figure*}
\centering
\epsscale{.17}
\plotone{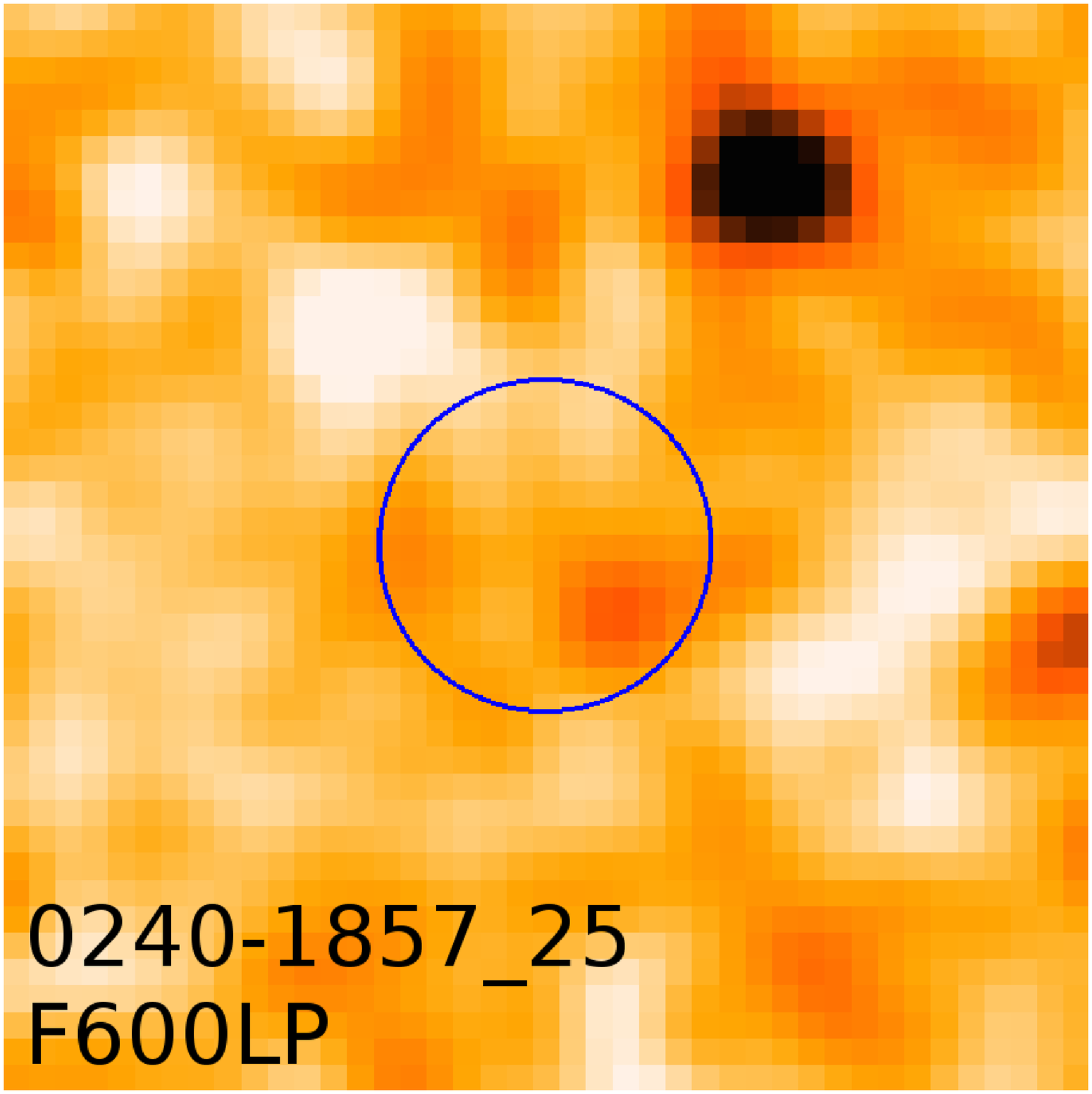}
\plotone{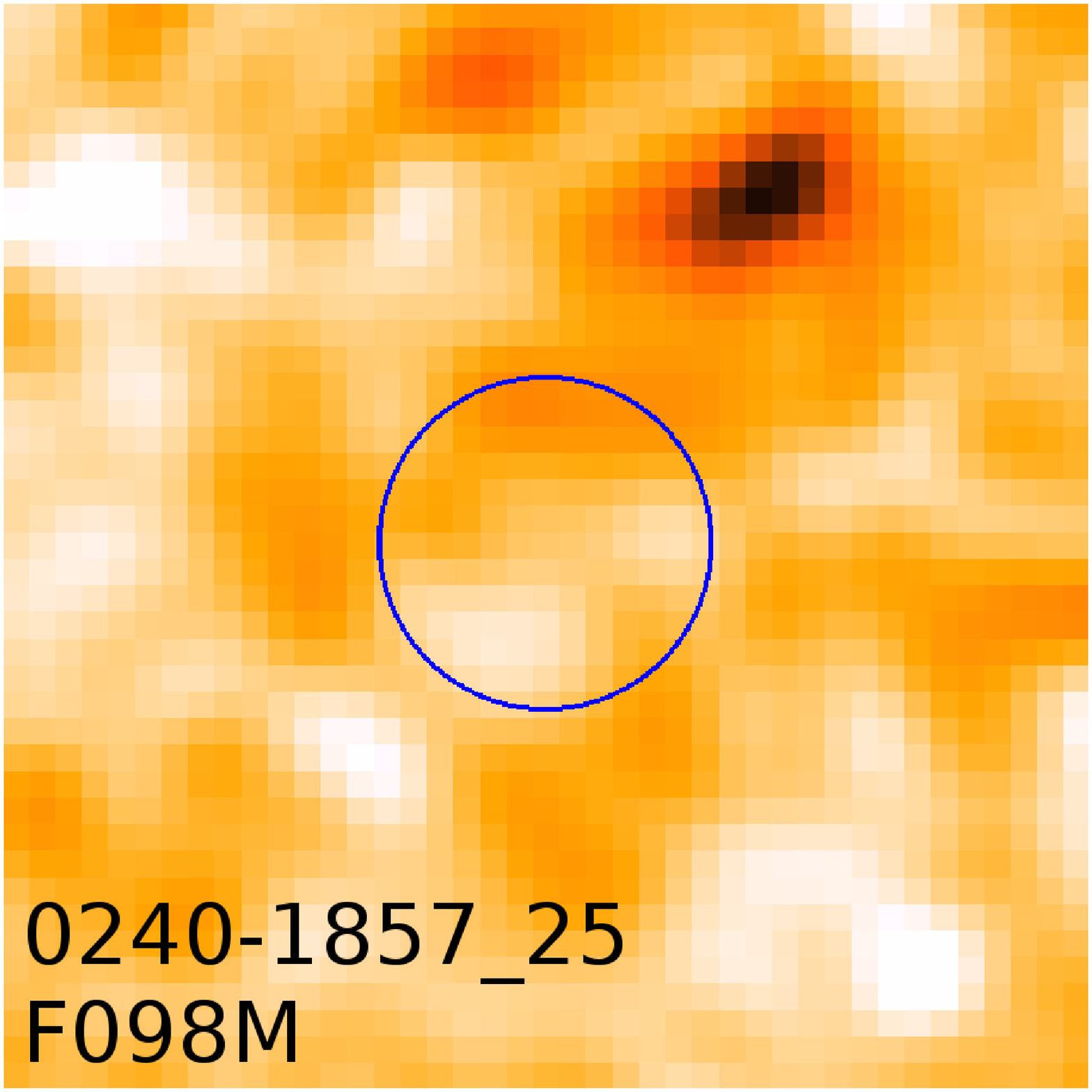}
\plotone{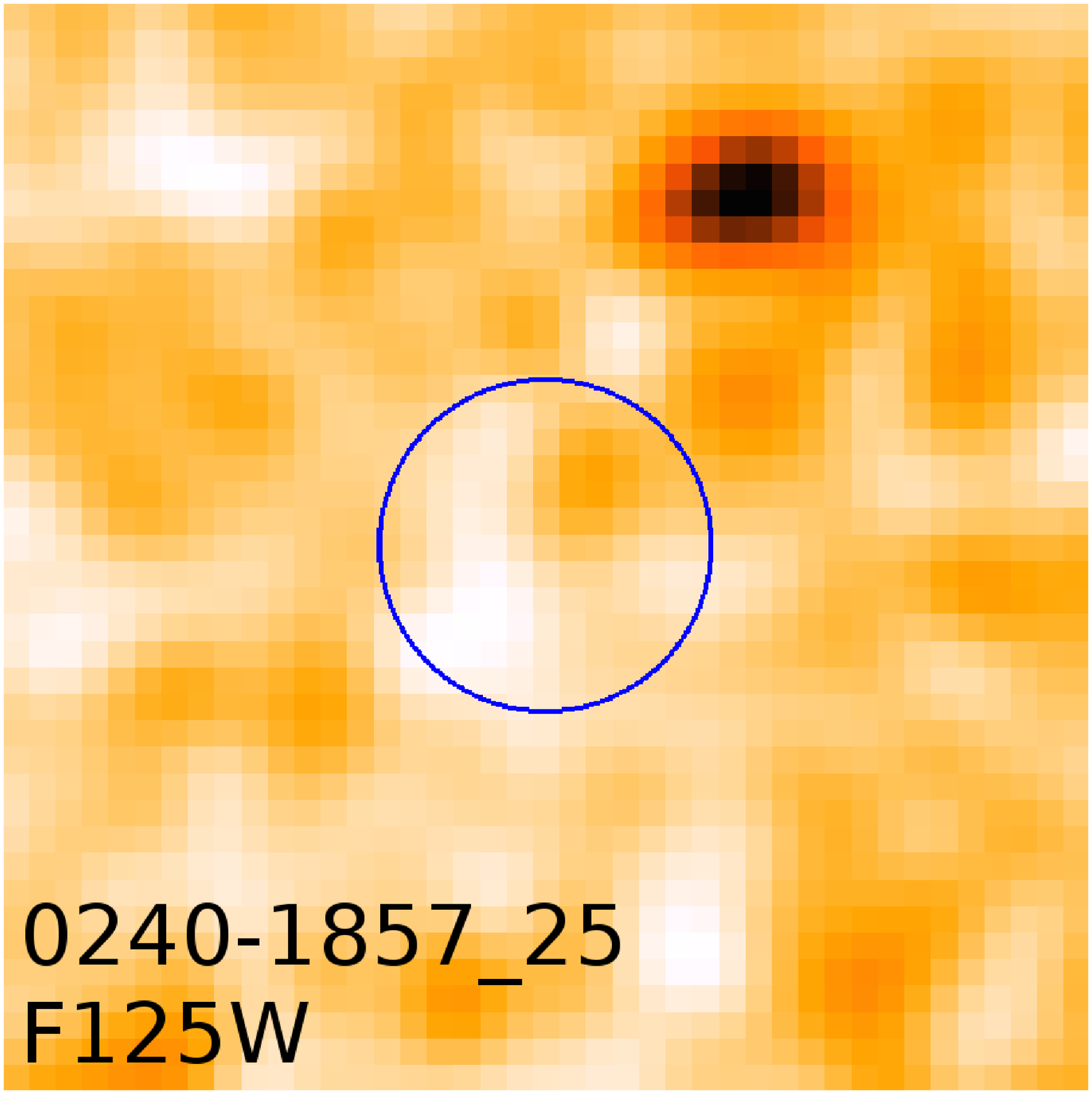}
\plotone{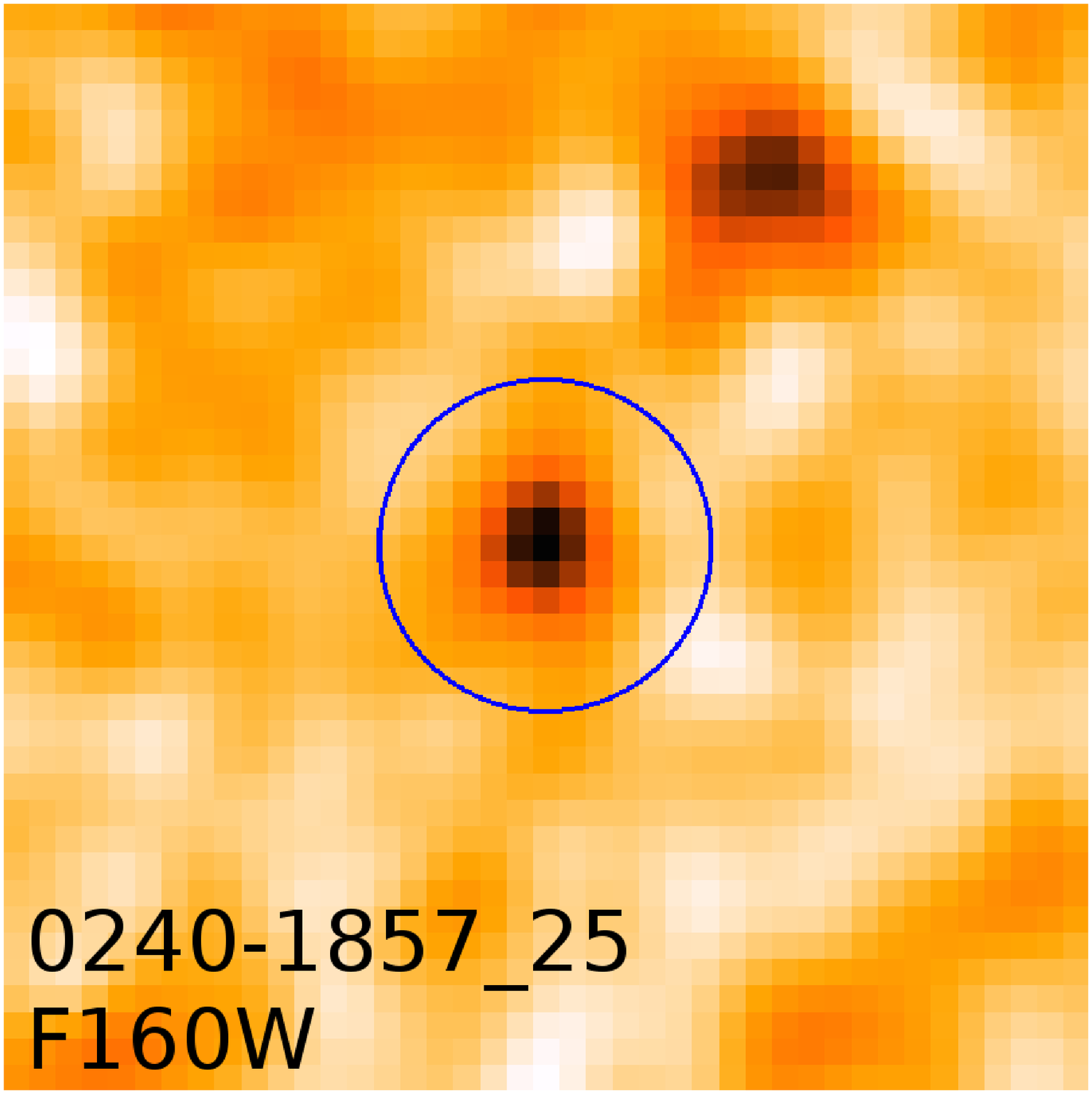}
\epsscale{.26}
\plotone{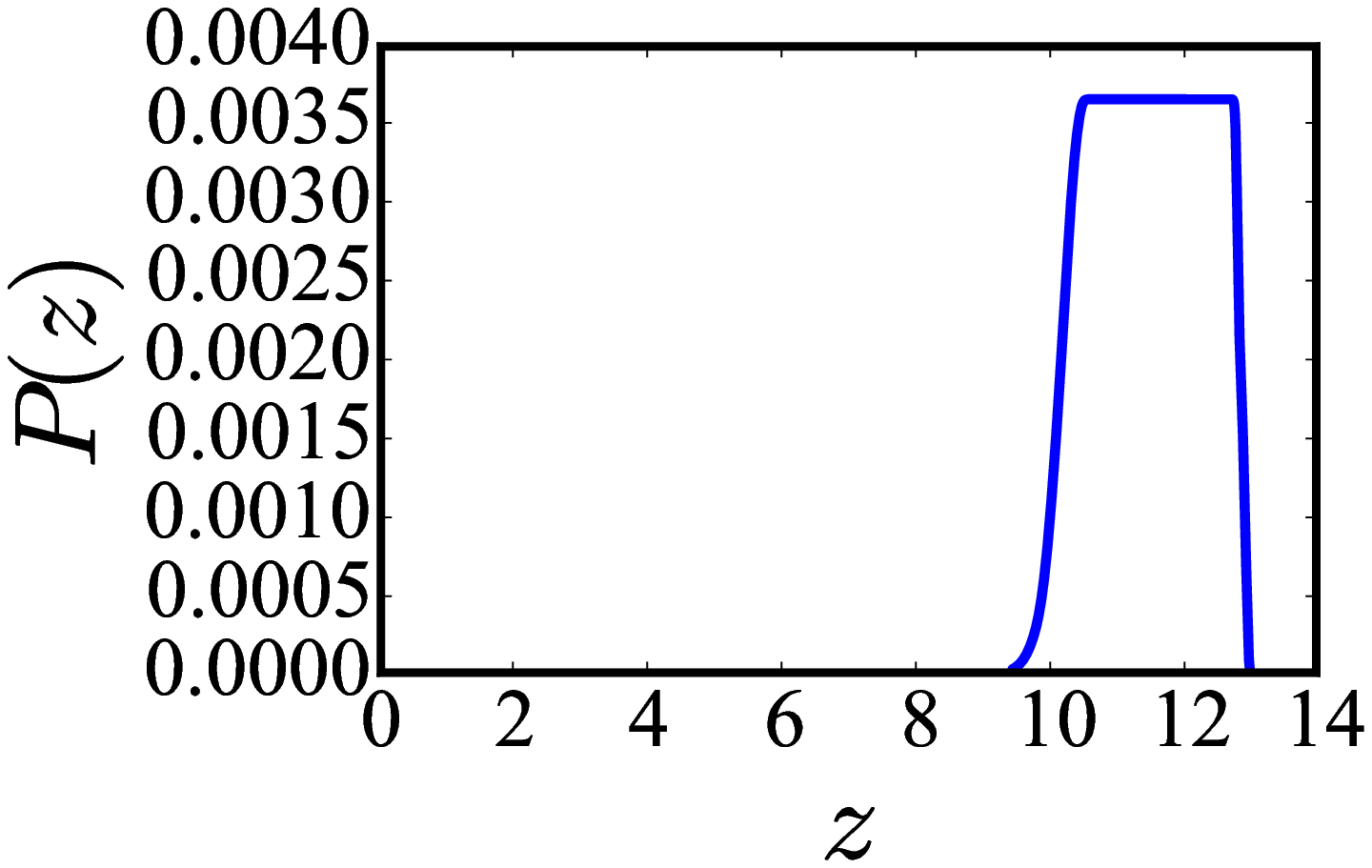}\\
\vspace{5pt}
\epsscale{.17}
\plotone{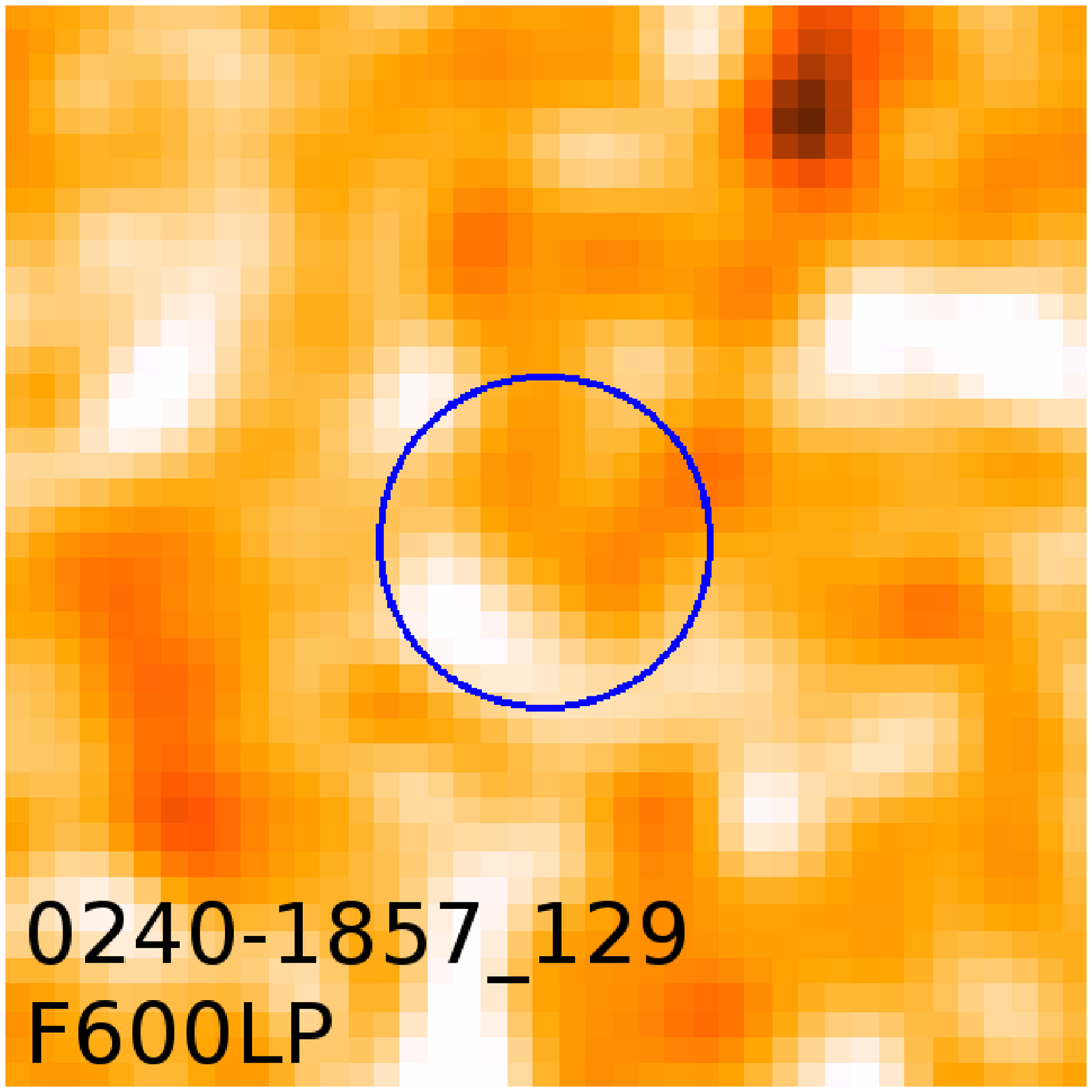}
\plotone{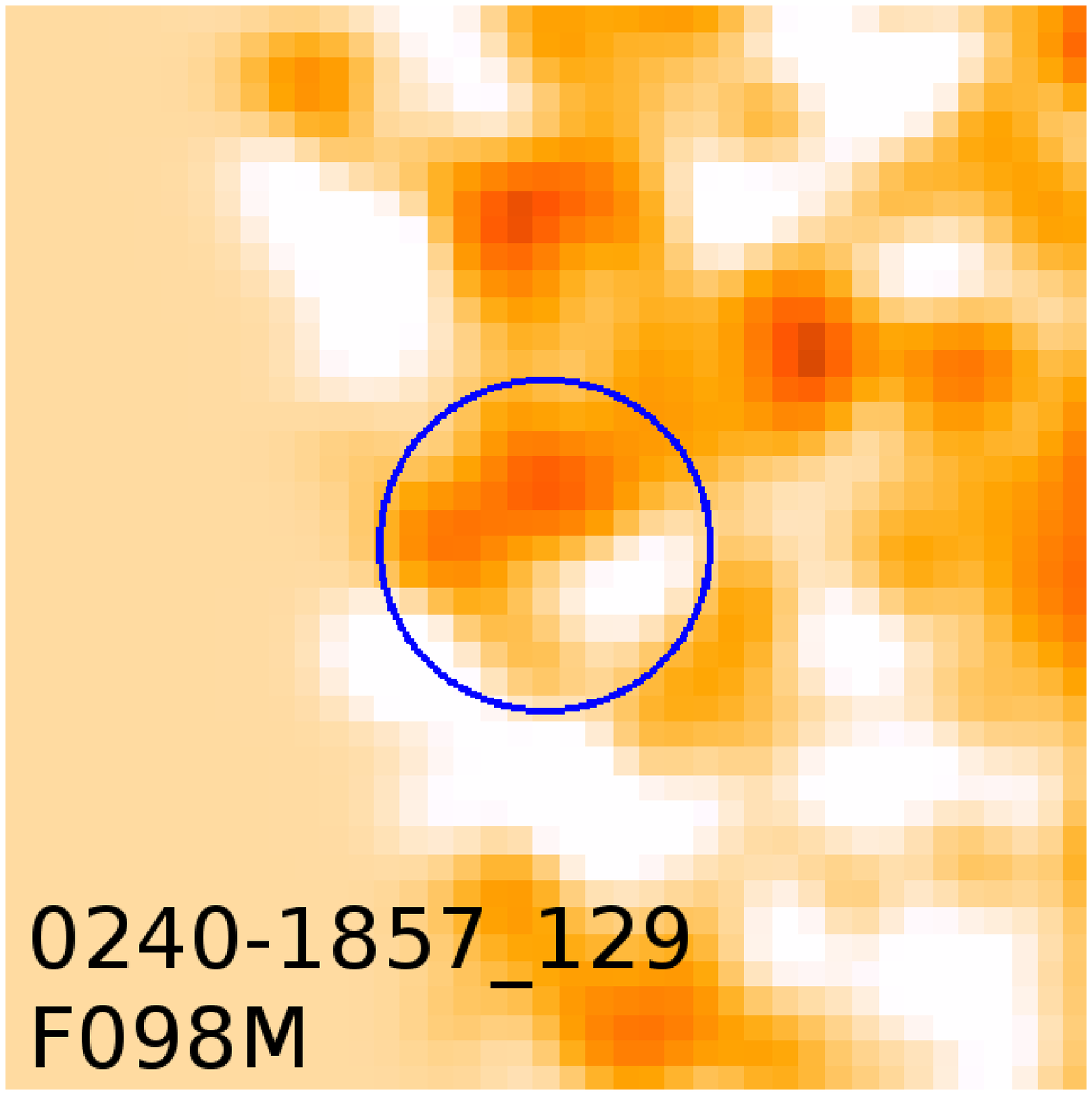}
\plotone{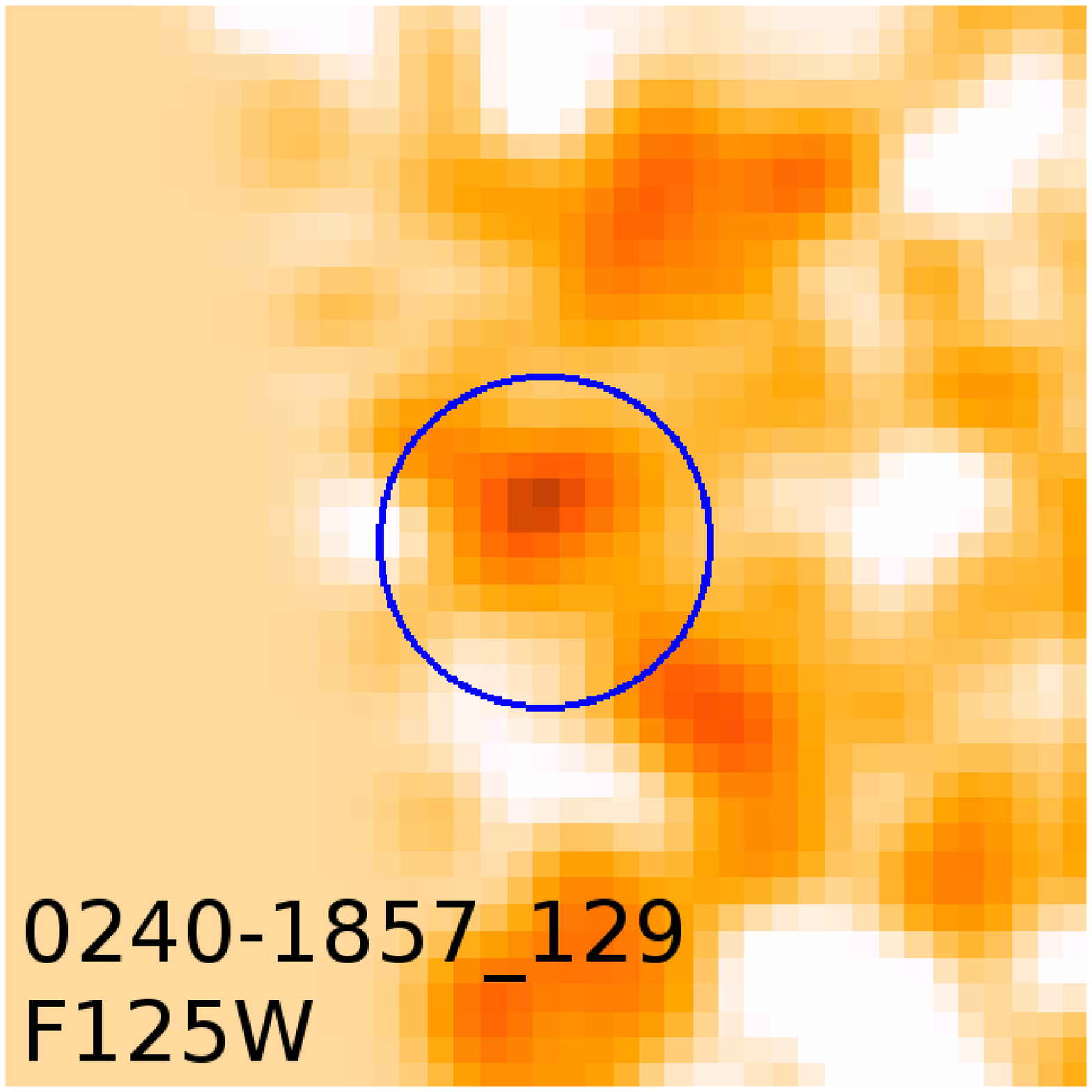}
\plotone{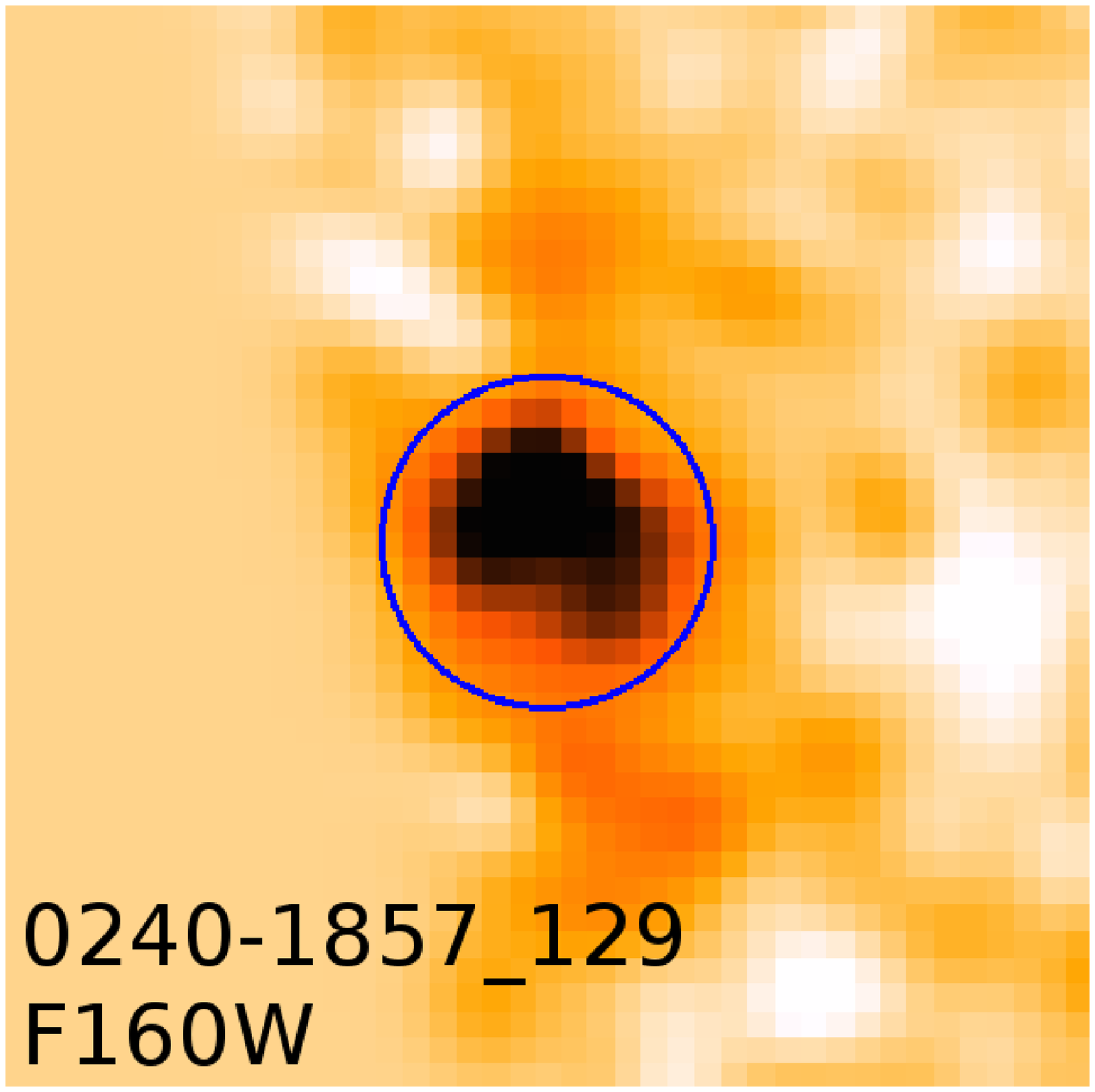}
\epsscale{.26}
\plotone{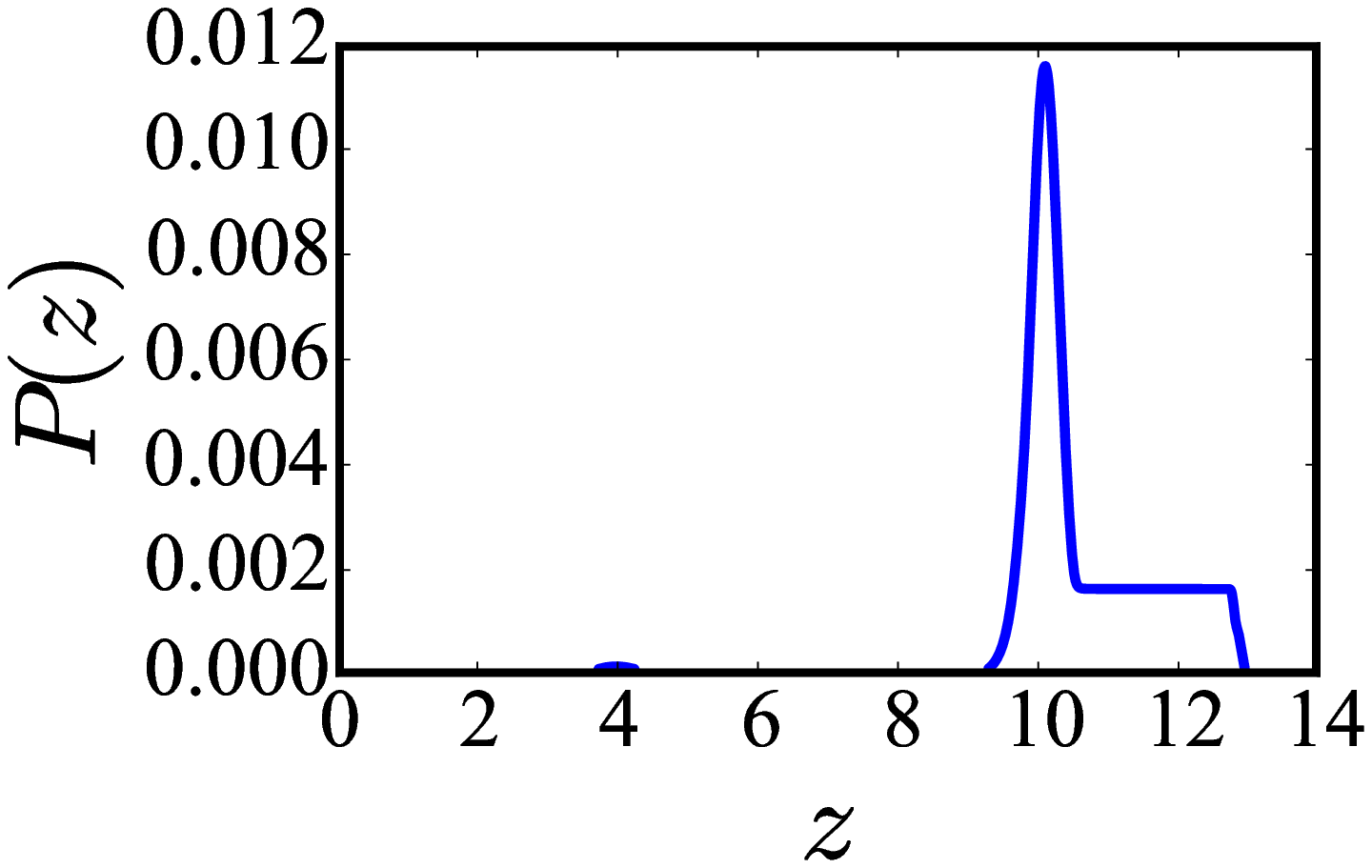}\\
\vspace{5pt}
\epsscale{.17}
\plotone{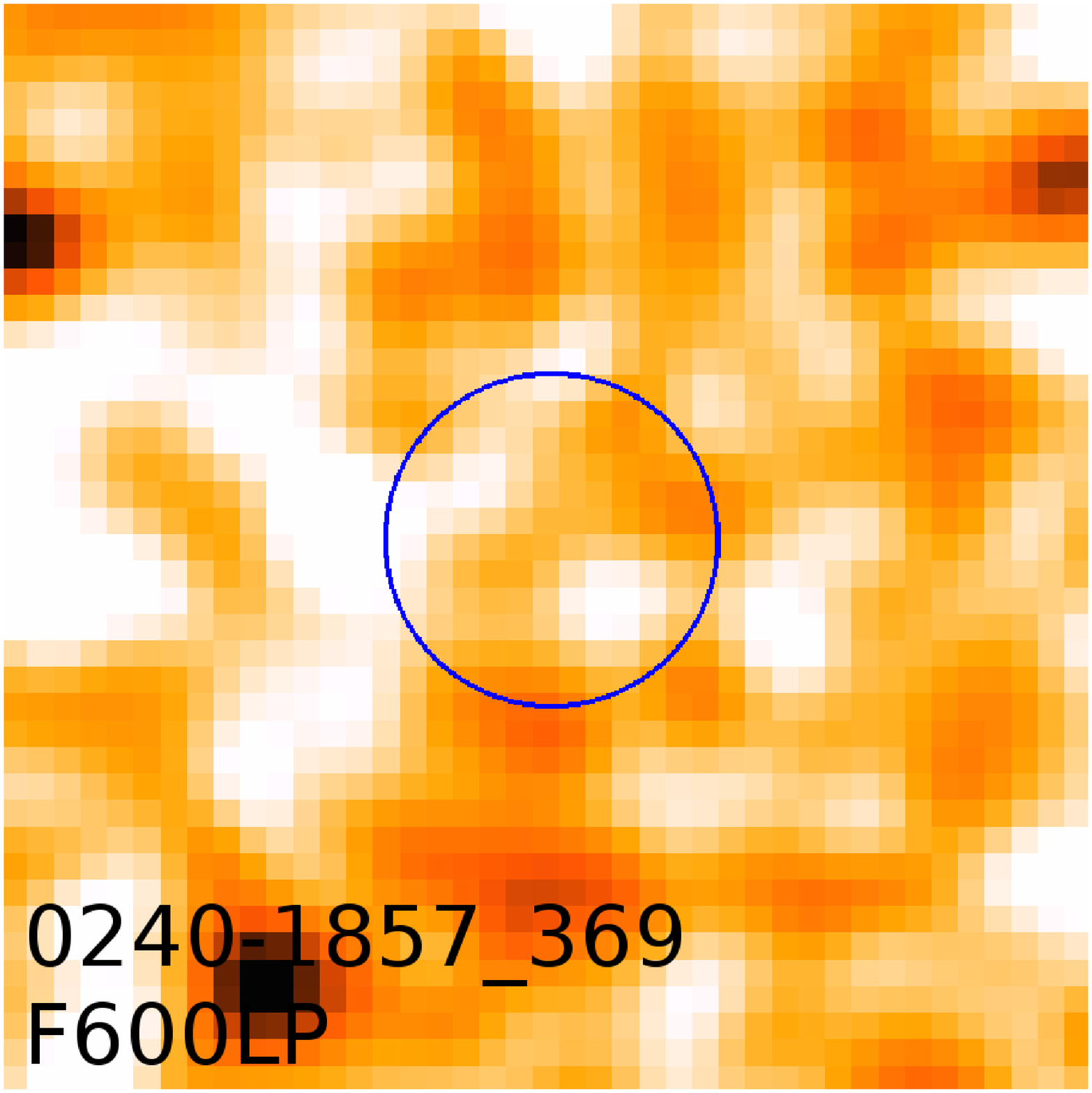}
\plotone{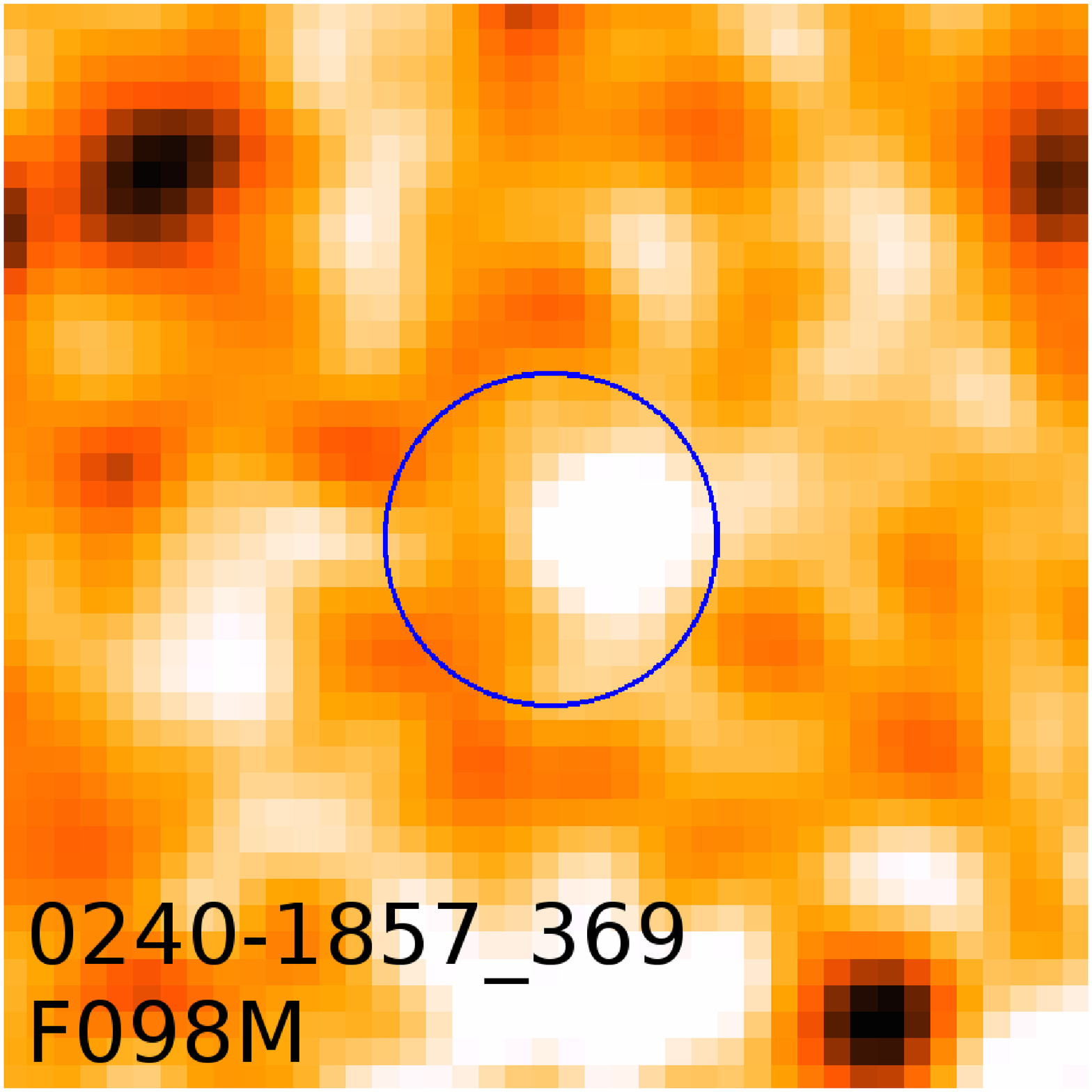}
\plotone{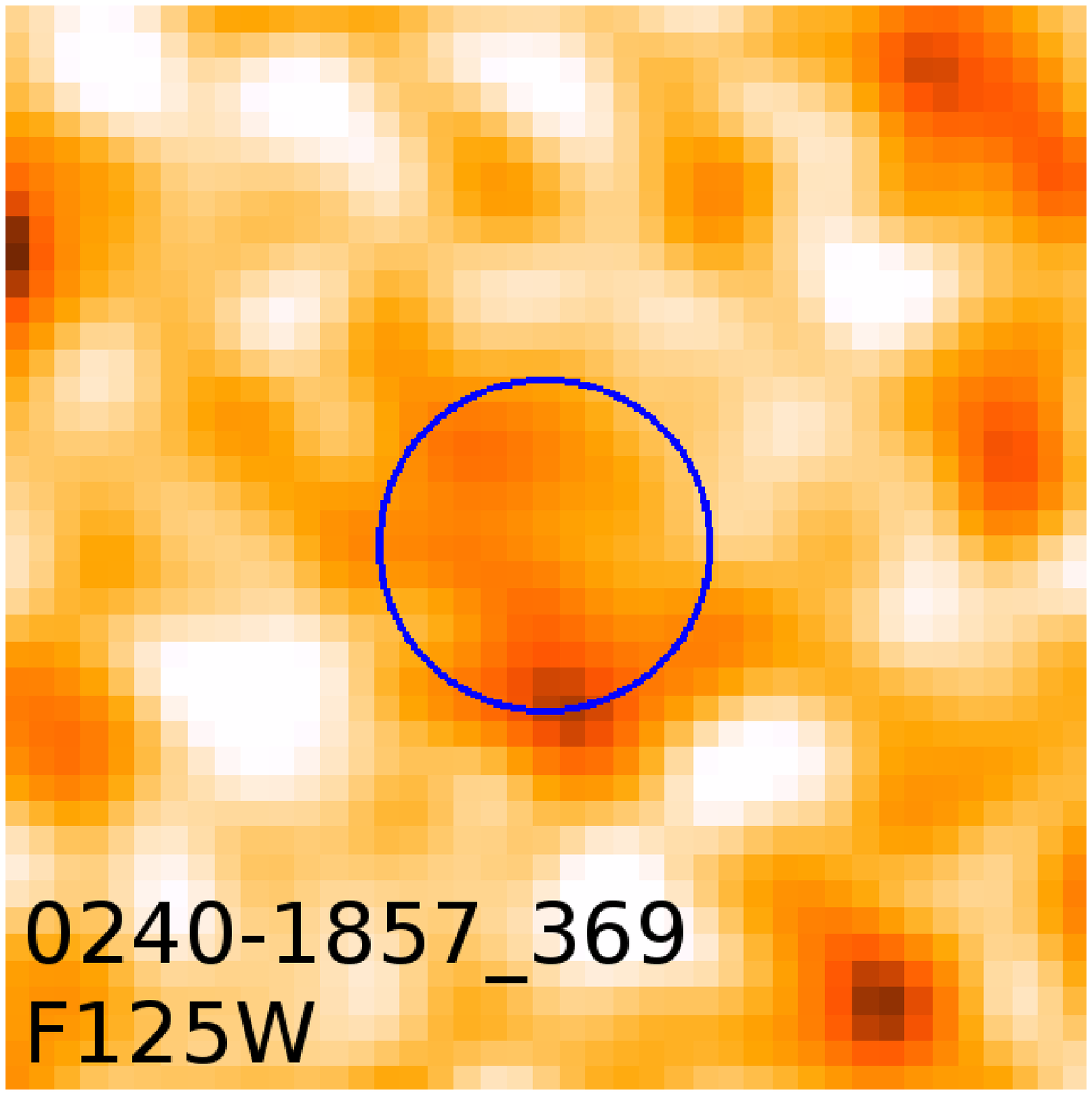}
\plotone{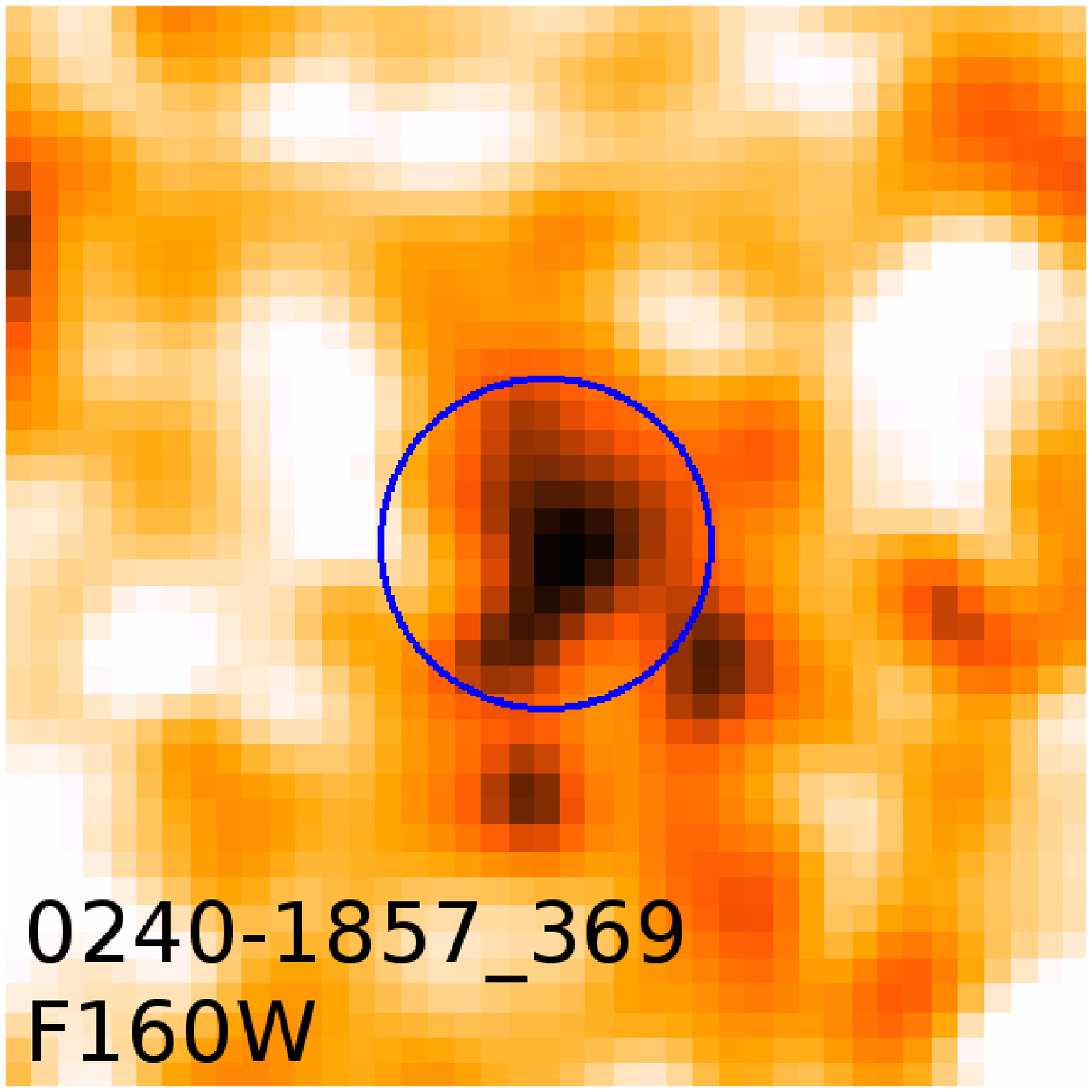}
\epsscale{.26}
\plotone{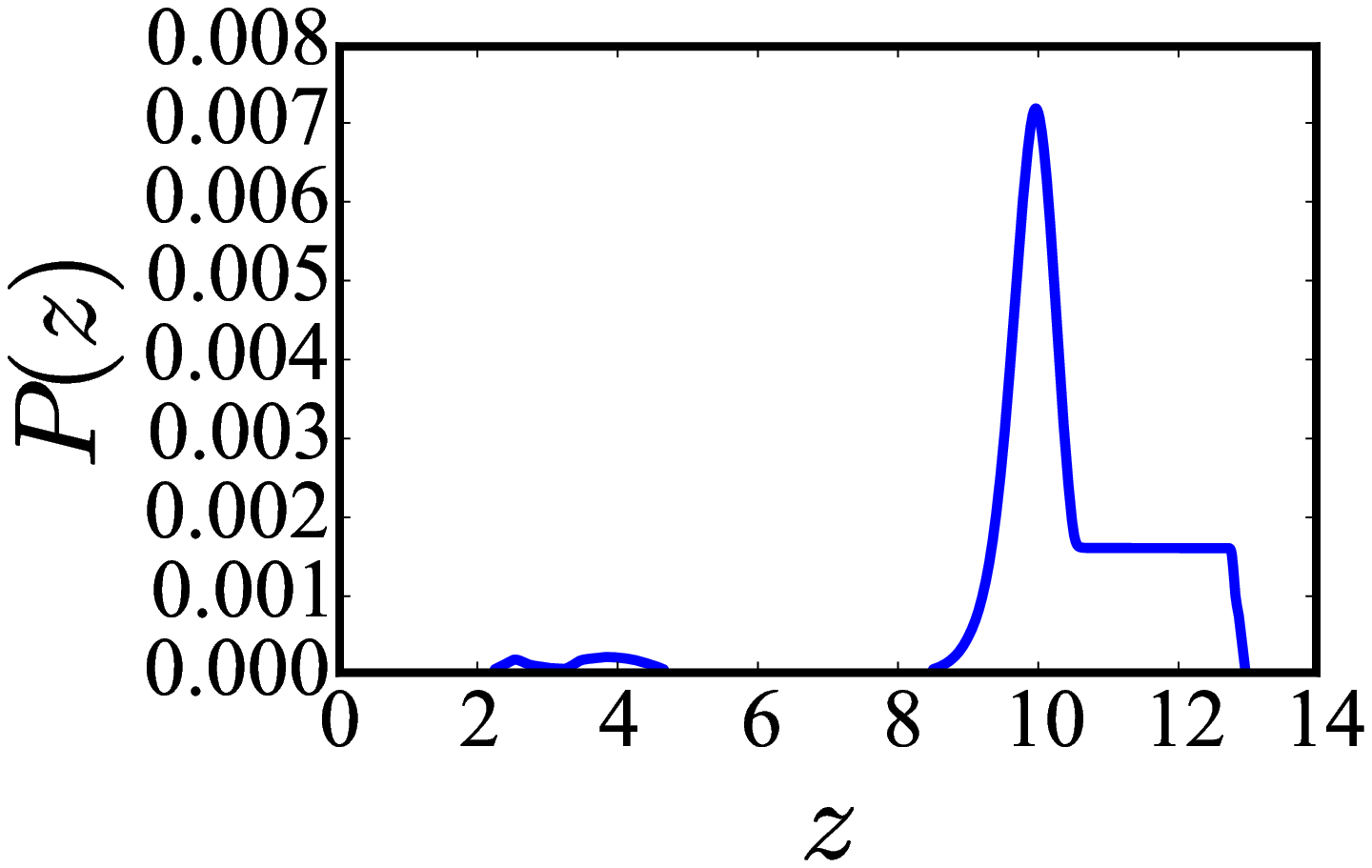}\\
\vspace{5pt}
\epsscale{.17}
\plotone{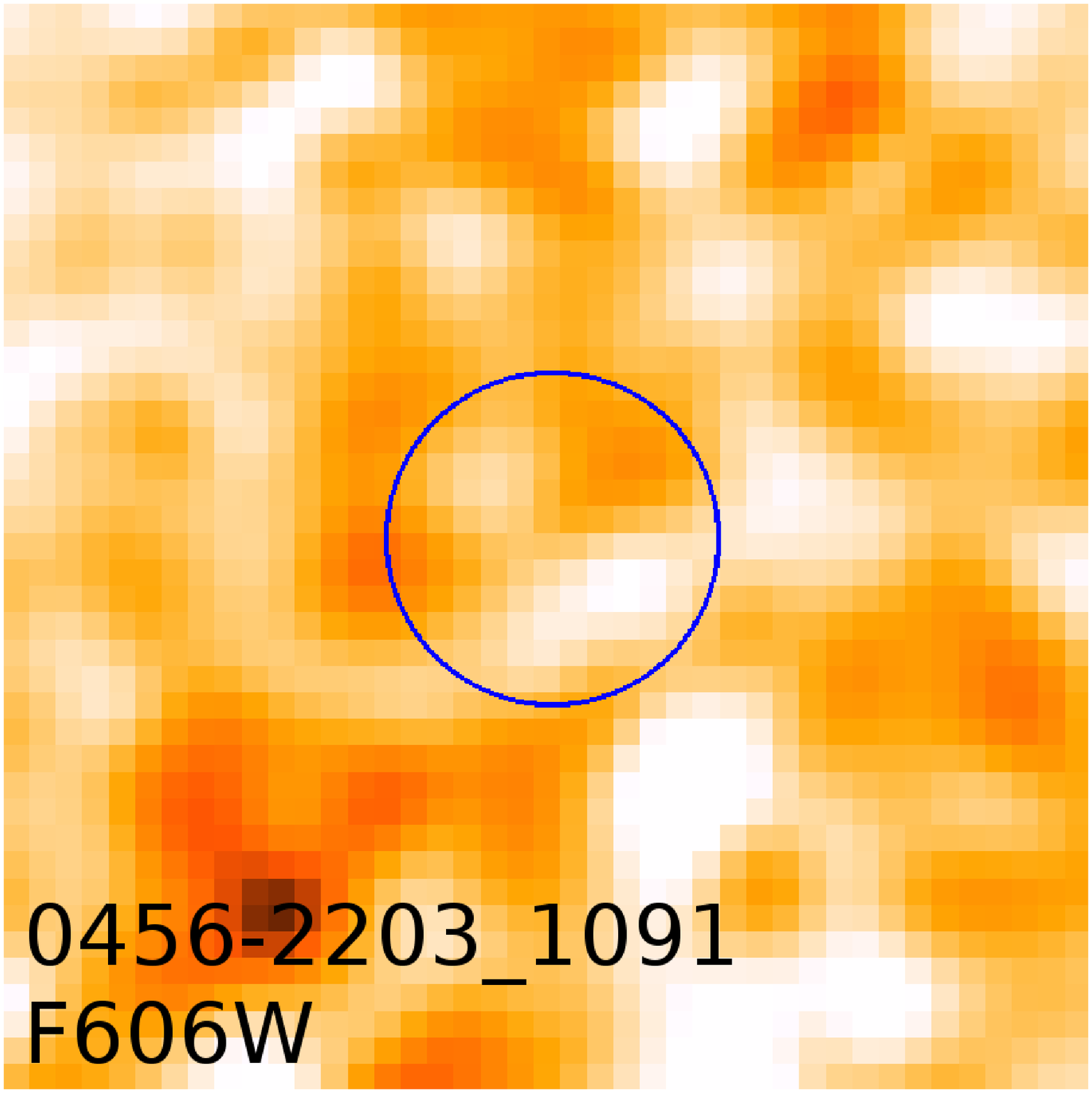}
\plotone{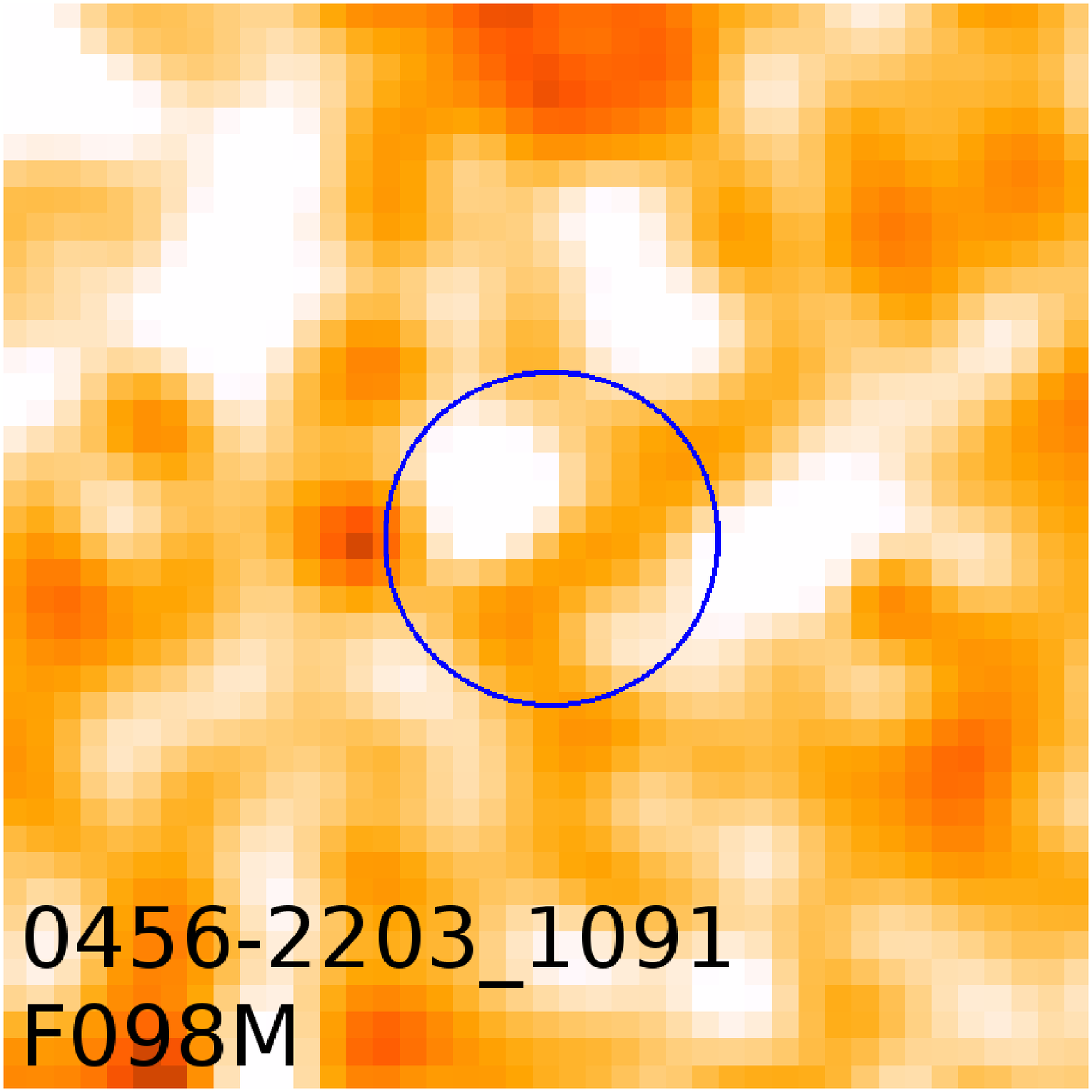}
\plotone{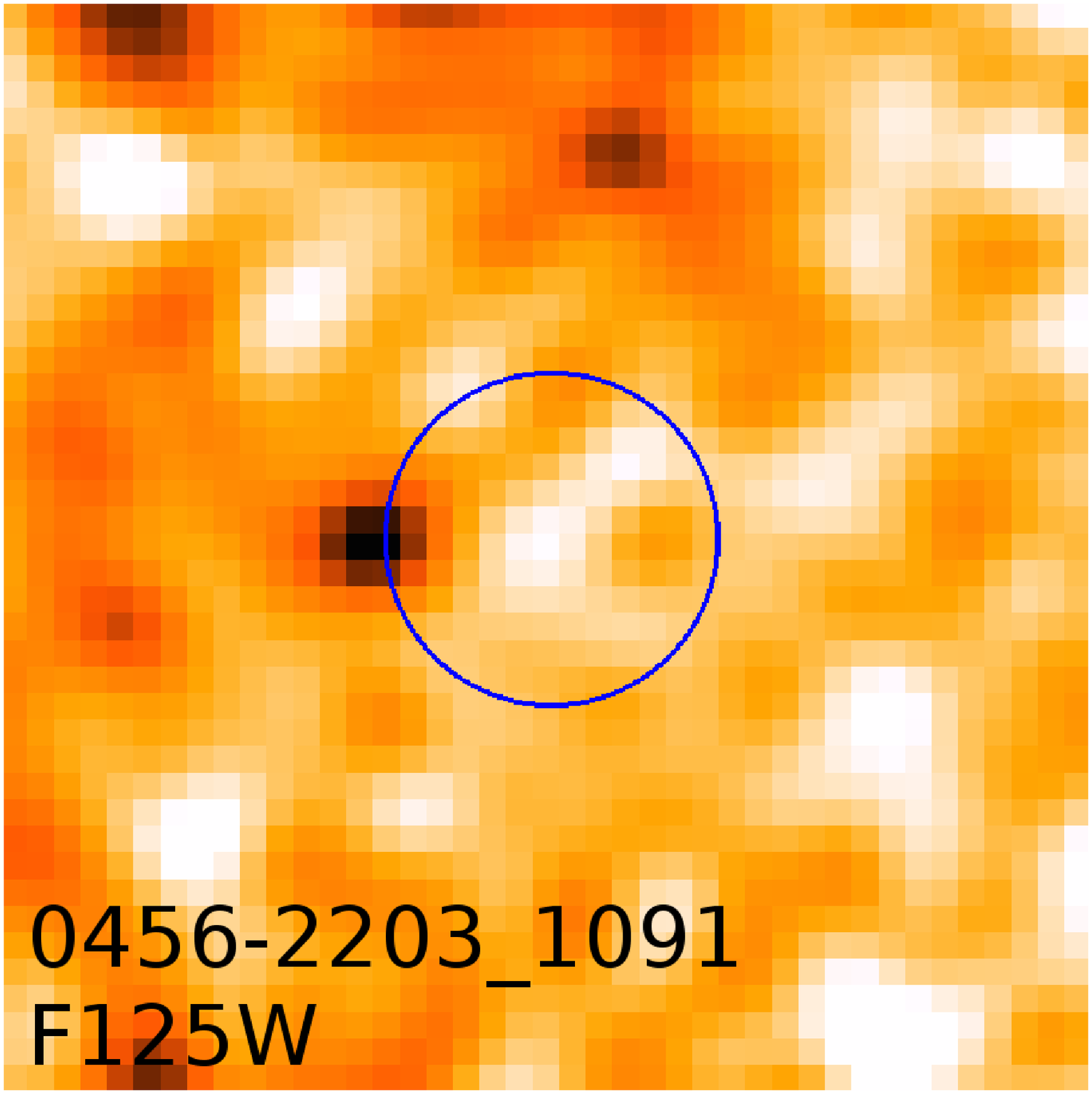}
\plotone{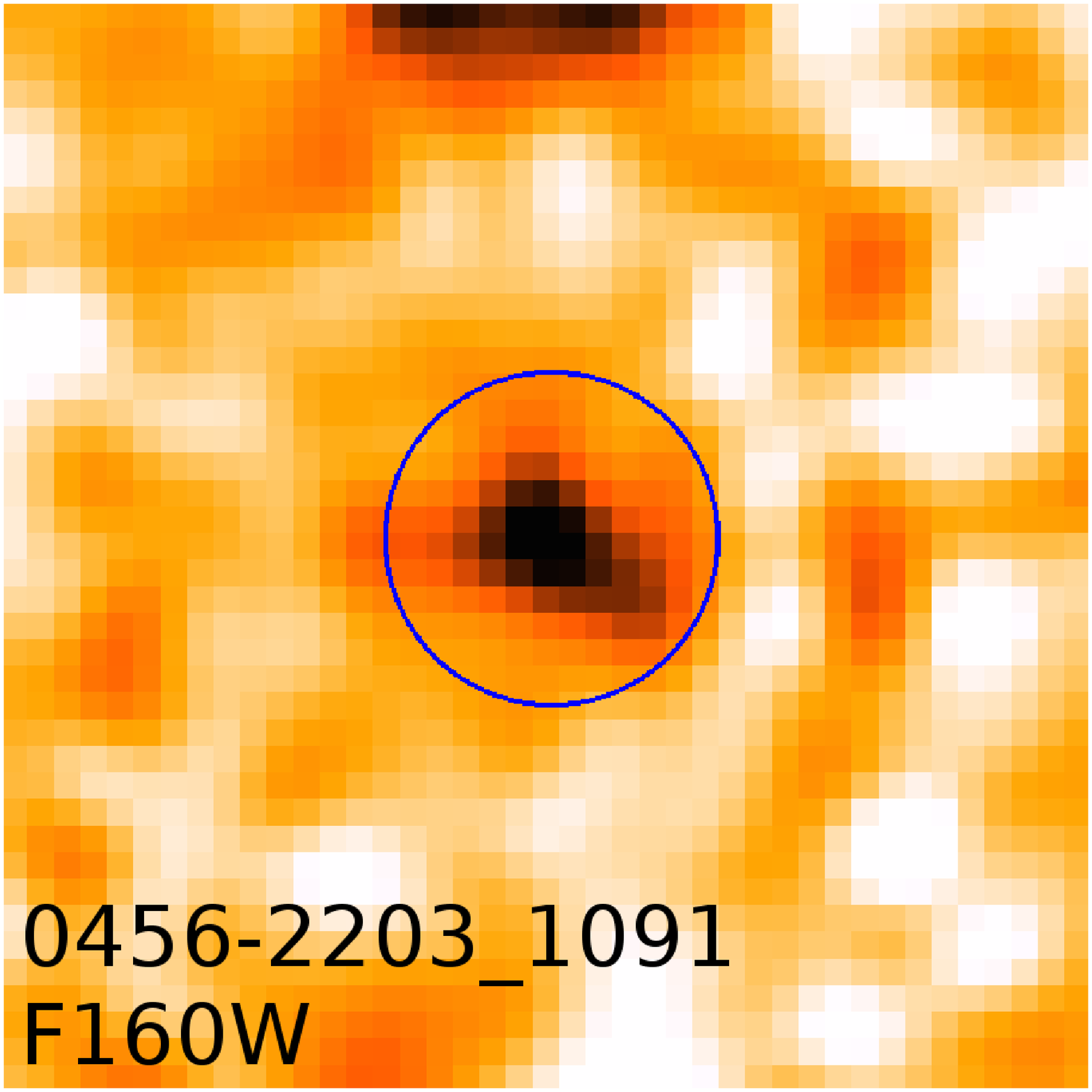}
\epsscale{.26}
\plotone{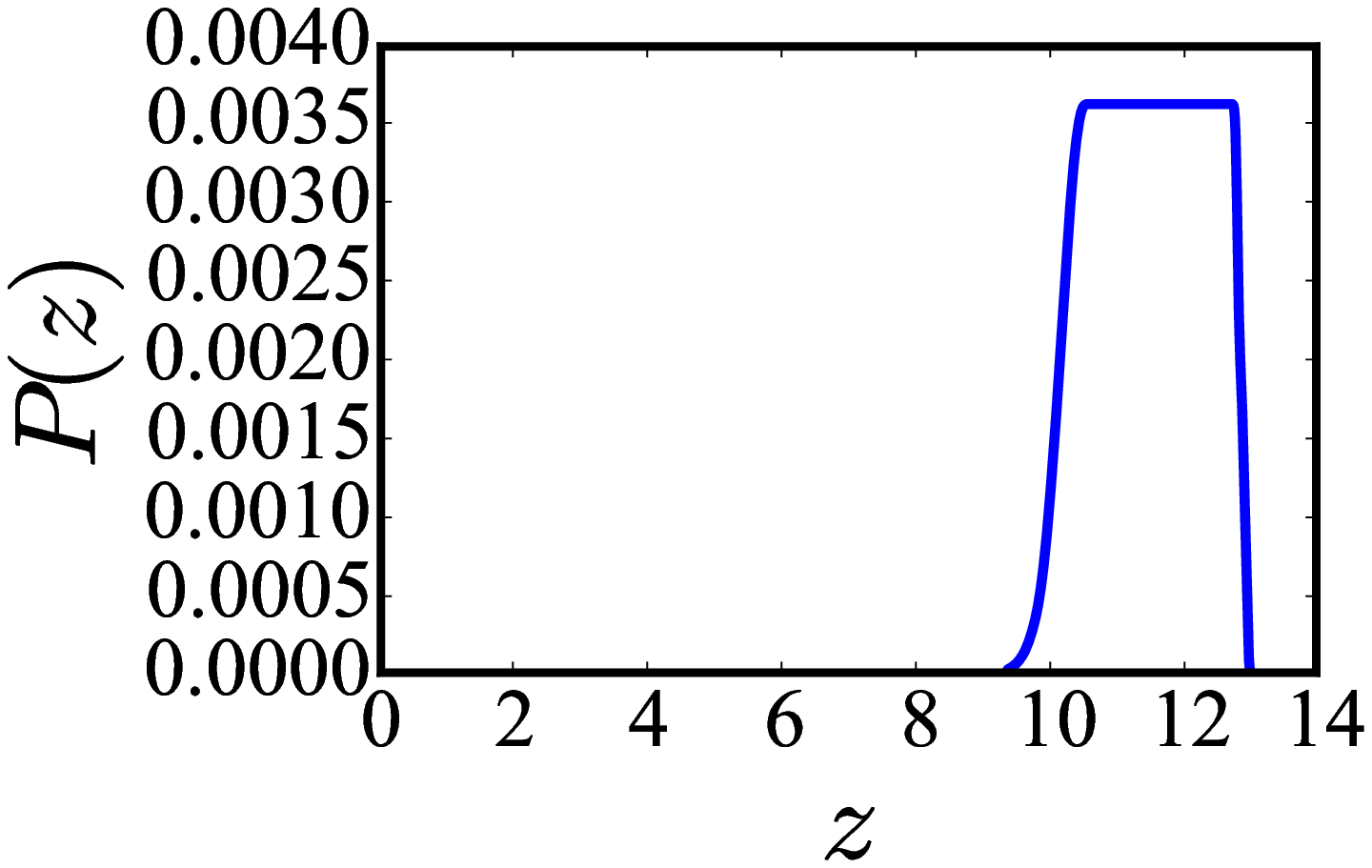}\\
\vspace{5pt}
\epsscale{.17}
\plotone{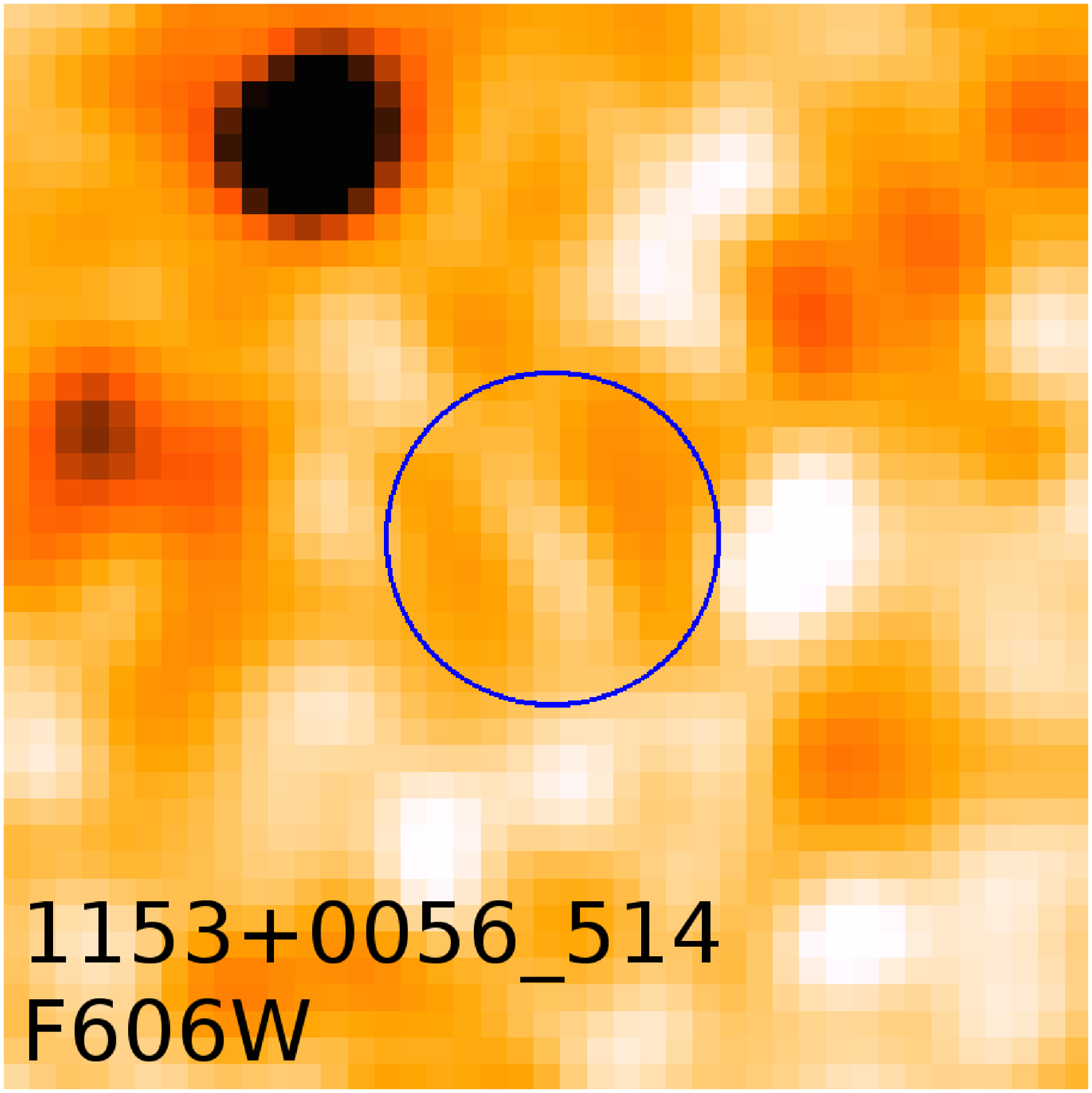}
\plotone{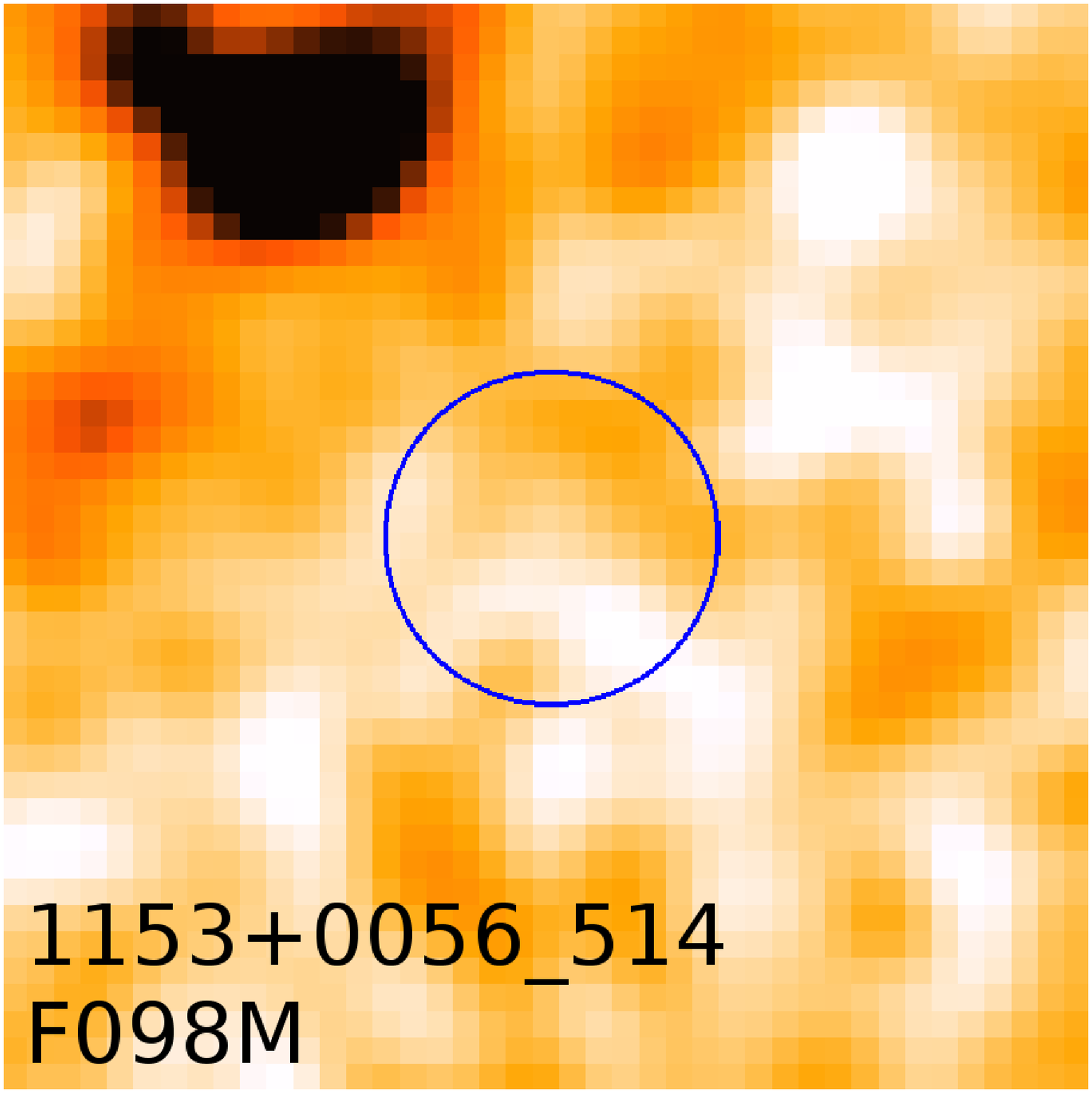}
\plotone{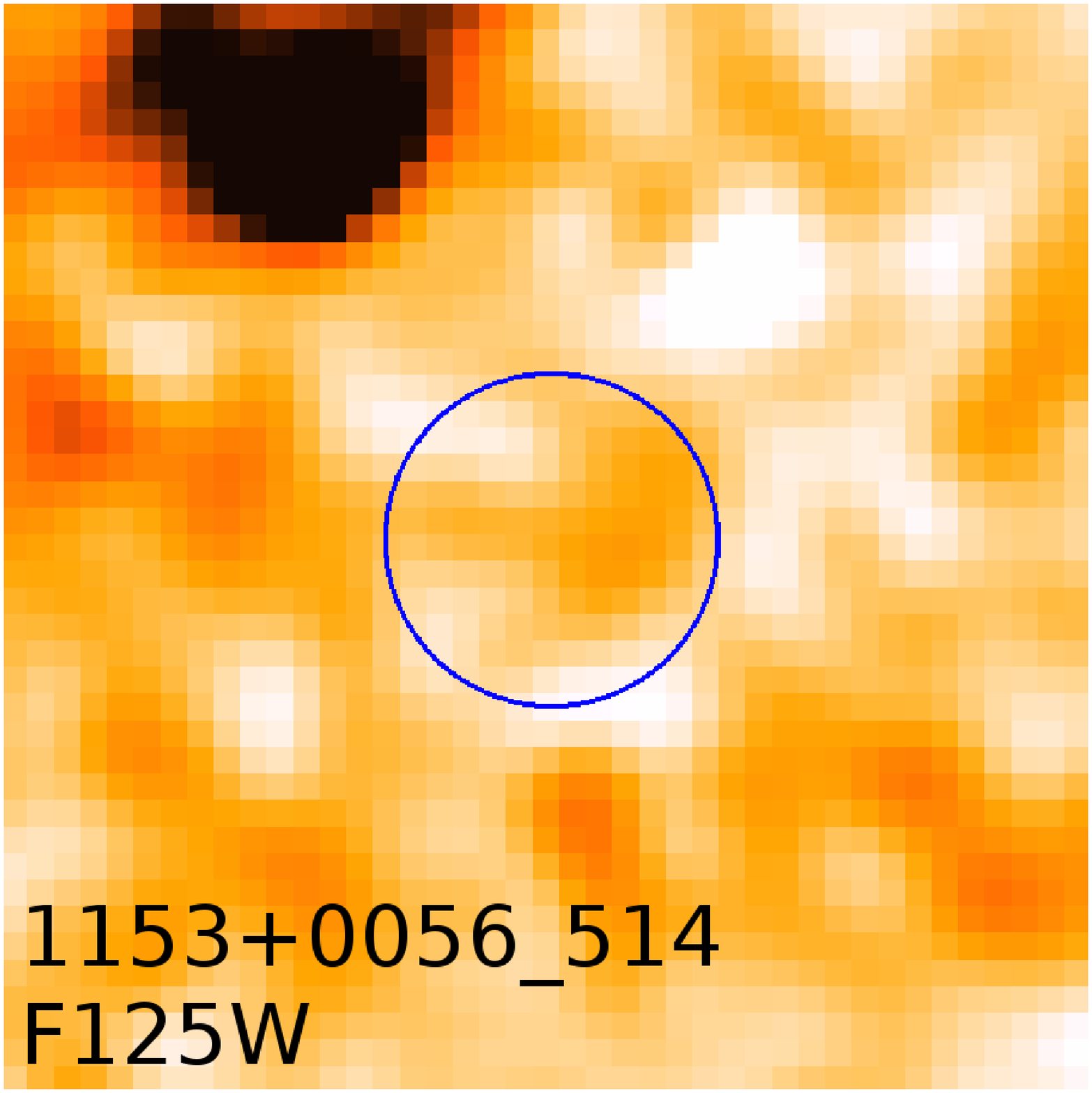}
\plotone{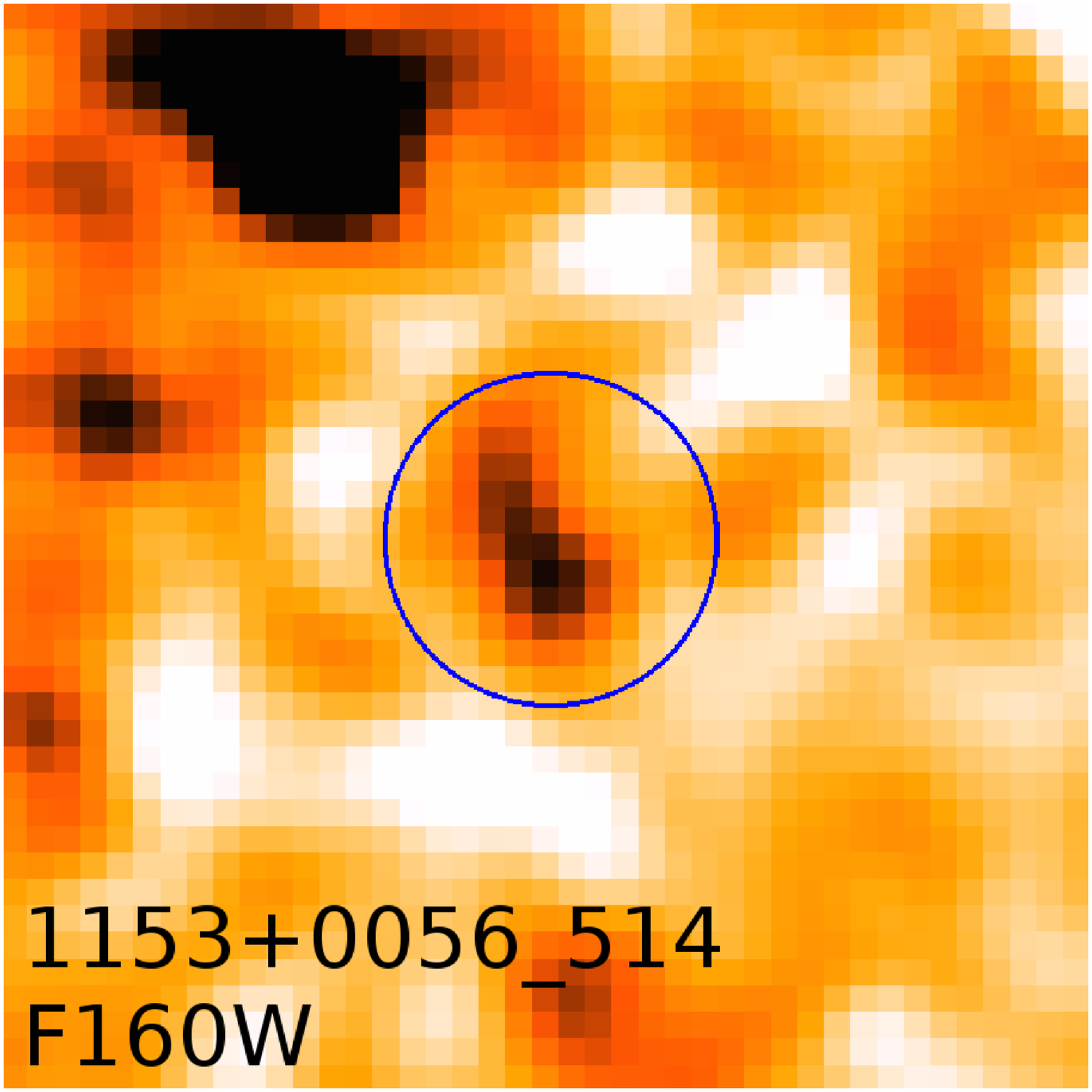}
\epsscale{.26}
\plotone{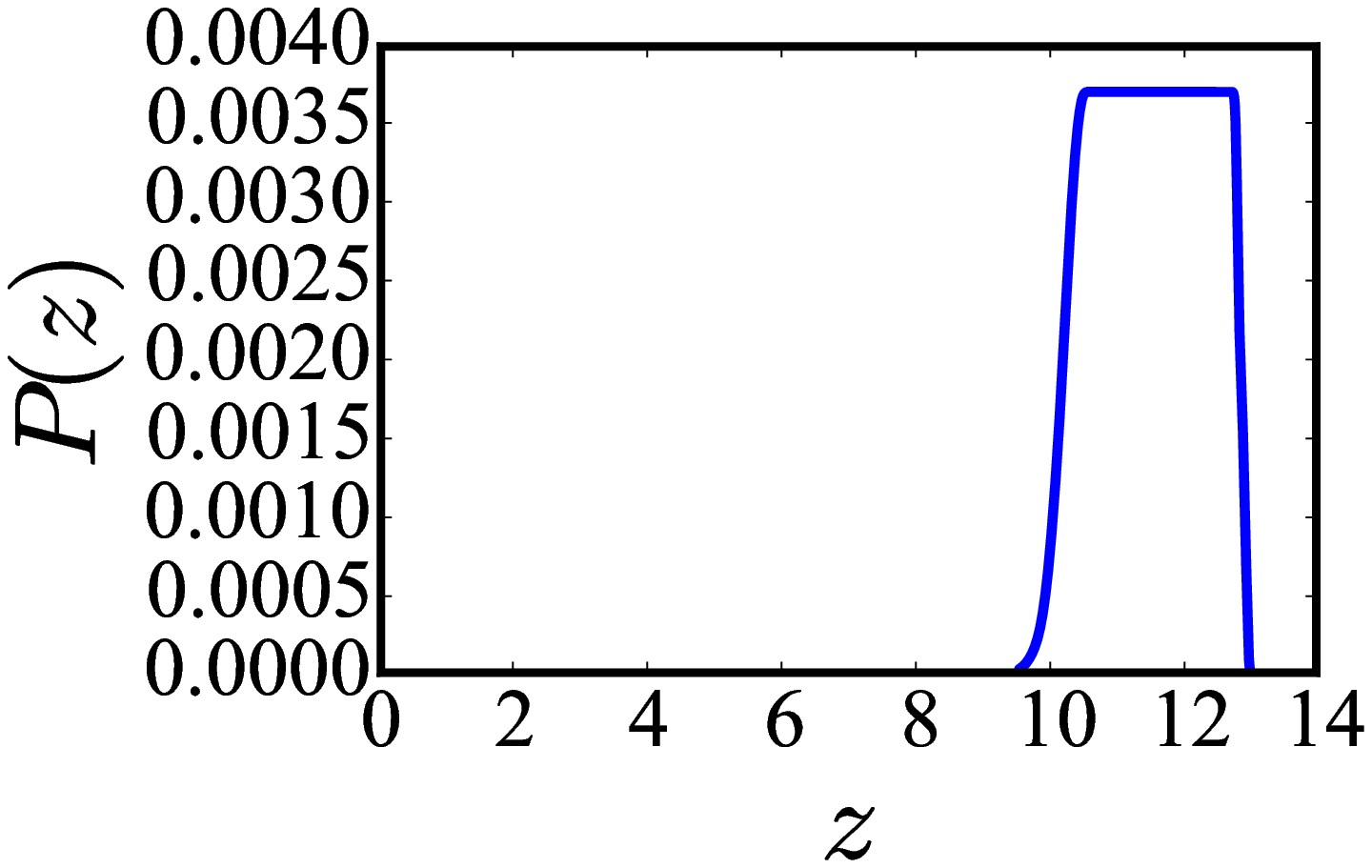}\\
\vspace{5pt}
\epsscale{.17}
\plotone{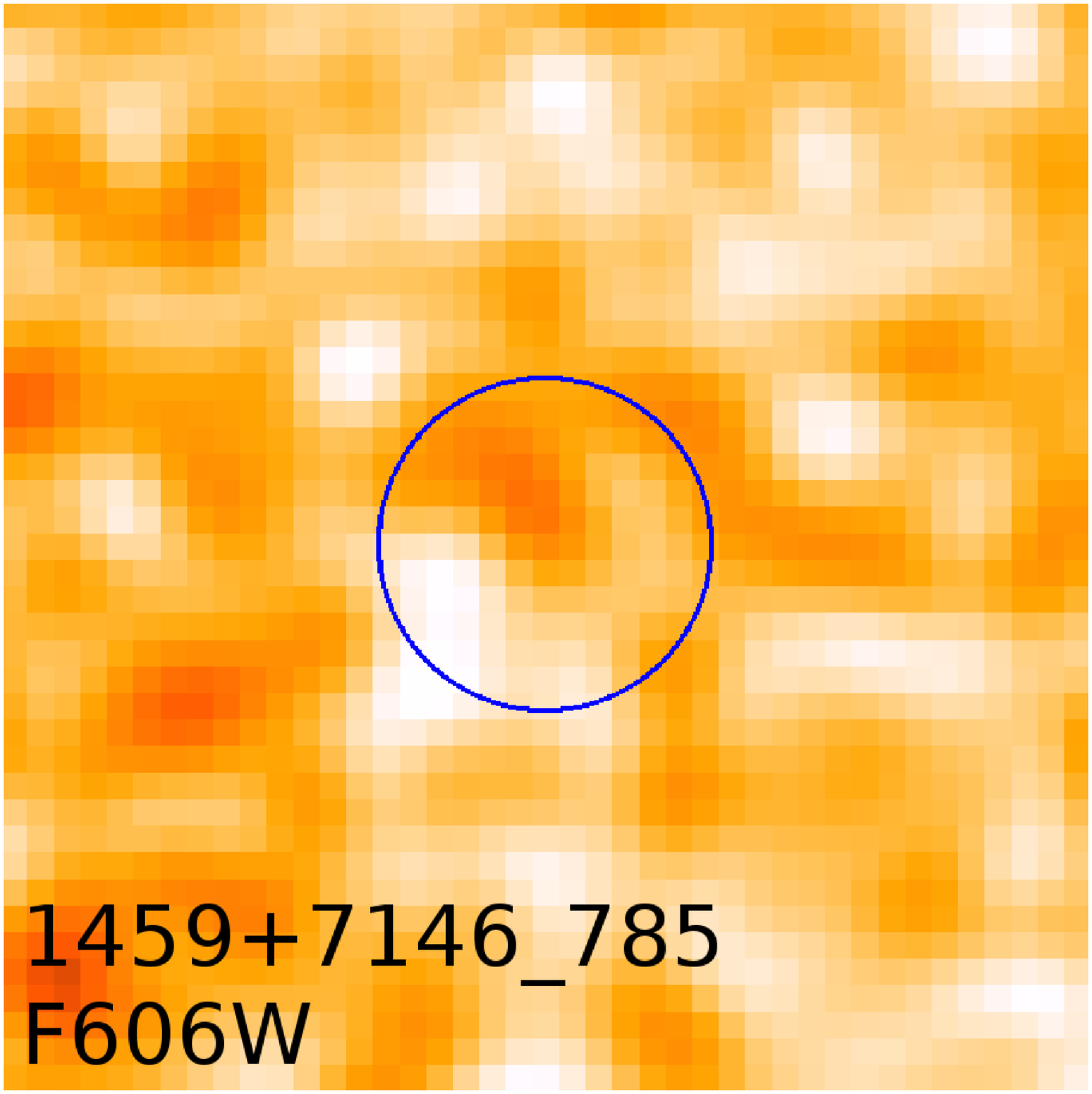}
\plotone{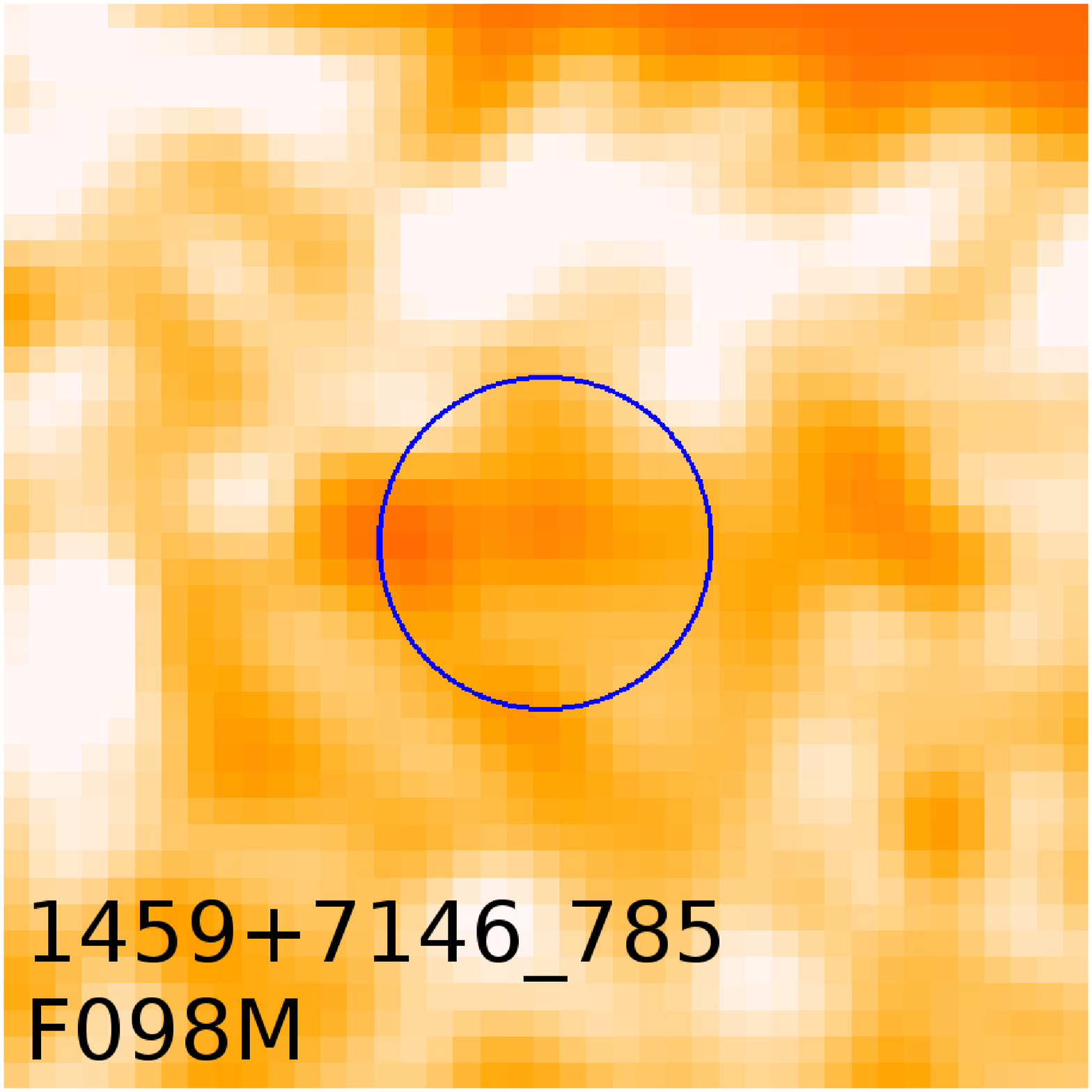}
\plotone{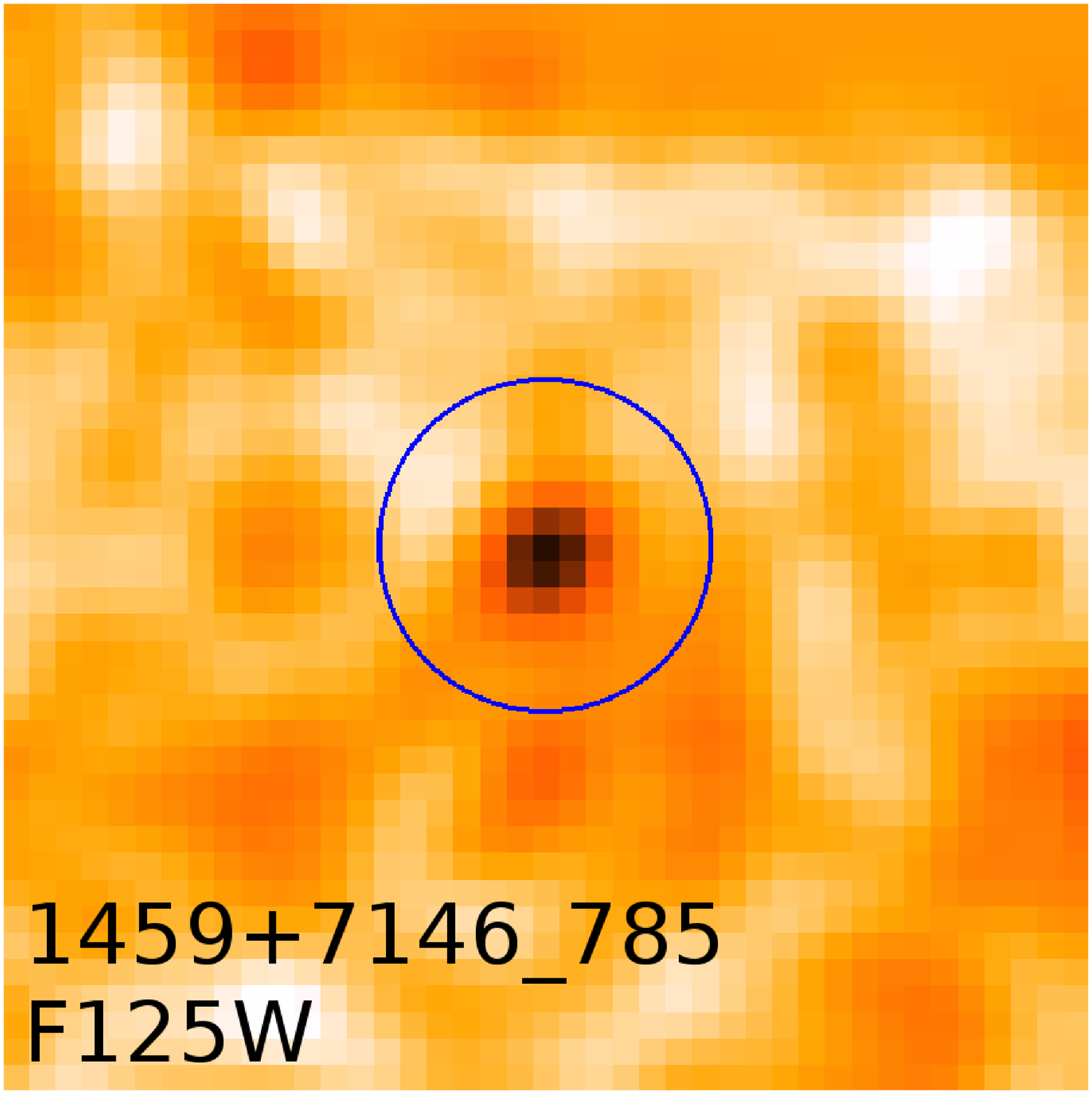}
\plotone{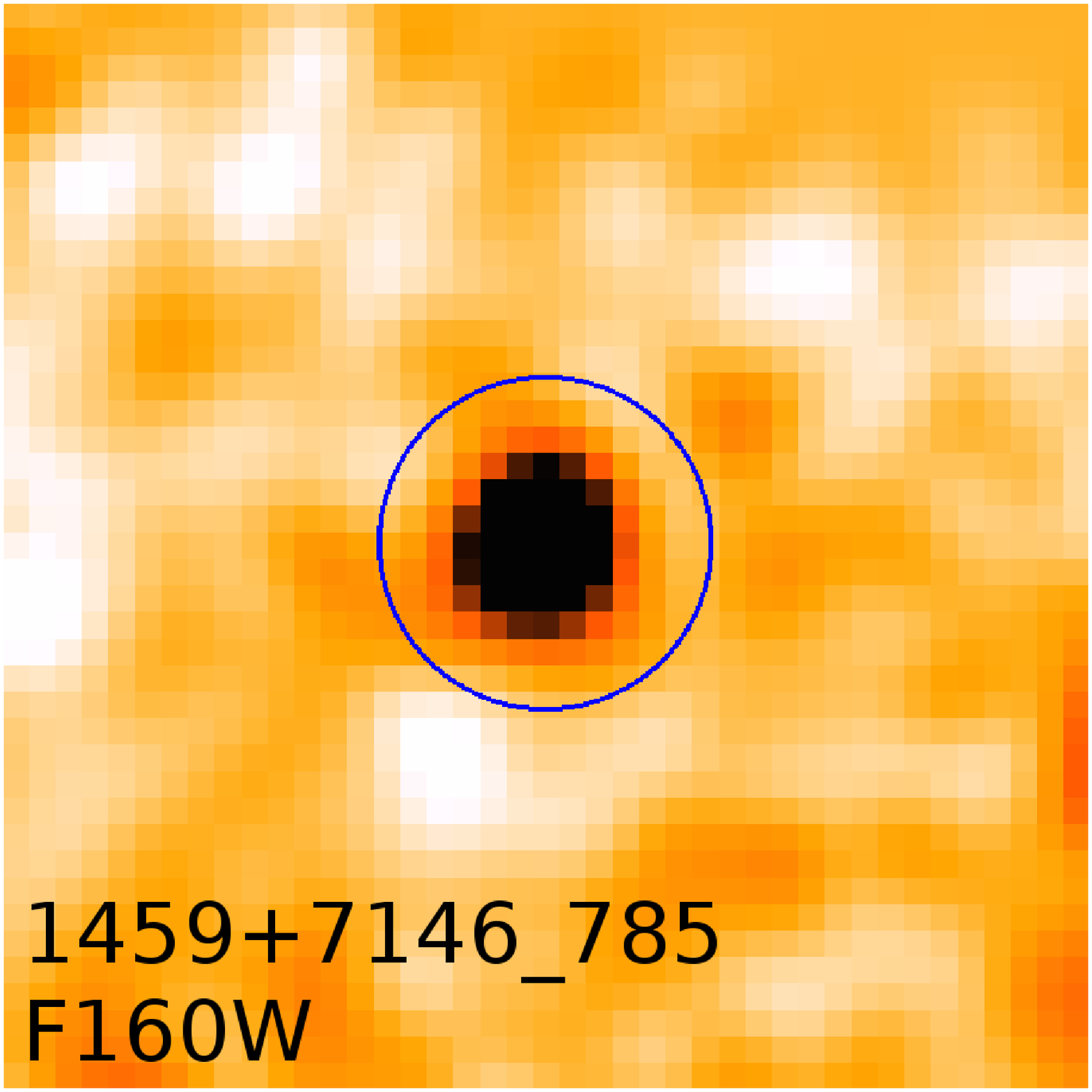}
\epsscale{.26}
\plotone{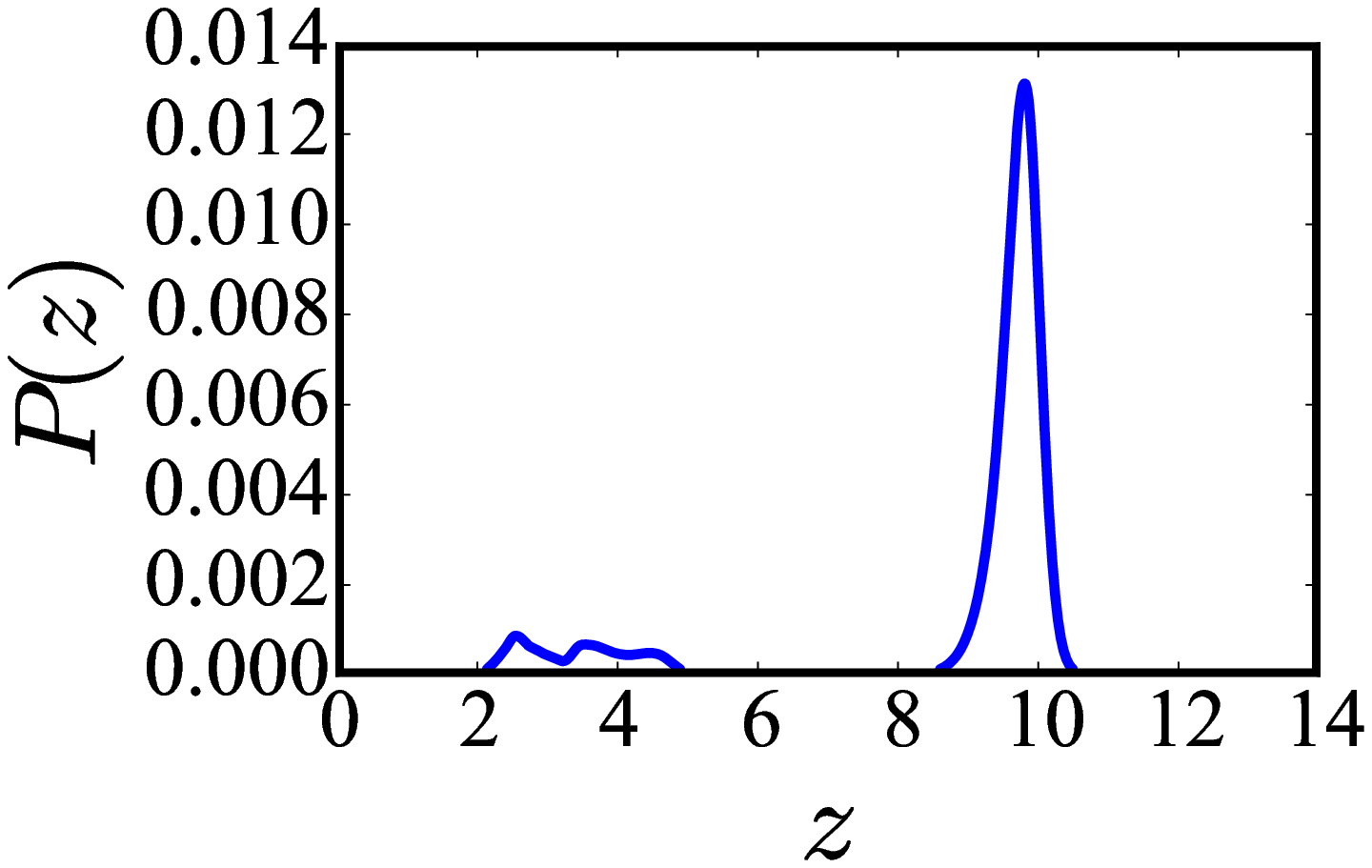}\\
\caption{Postage stamps of the \jband-band dropouts listed in Table \ref{table1}. Each image is $3\farcs2 \times 3\farcs2$. The diameter of each circle is $1\farcs0$. Each image and circle is centred on the candidate dropout galaxy. The left four columns show the candidate in $V$, \yband, \jband{}, and \hband{}, while the right-most column shows the redshift probability distribution $P(z)$ vs $z$ determined from BPZ for each candidate. \label{fig1}}
\end{figure*}

We performed a search for \jband-dropouts over $293$ arcmin$^2$ of
archival BoRG data. We find six sources that satisfy
the \jband-dropout selection with S/N$_H \ge 8$. The candidates are
detected over a range of magnitudes, with four candidates between
$H_{160} = 25.8-26.4$, and two brighter candidates at $H_{160} =24.7$
and $H_{160} =25.2$. At $z =10$, this corresponds to
$\mathbf{M}_{\mathrm{AB}} = -21.1$ to $-22.8$. Three candidates are detected
only in \hband{}, while the remaining three are detected in both
\jband{} and \hband{}.  The photometry of the candidates is reported
in Table \ref{table1}, and postage stamps of \vband, \yband, \jband{},
and \hband{} are shown in Figure \ref{fig1}.

We derive photometric redshifts for these six candidates using
the photo-$z$ code BPZ \citep{benitez00}, assuming a flat
prior on redshift, motivated by the uncertainty in the density of
sources at intermediate redshifts with colors similar to those of
$z\gtrsim 9$ galaxies. For the single band (\hband) detections, the
photometric redshift distribution is flat over the range $z \sim 10 - 13$. For the
two-band (\jband{} and \hband) detections, the photometric redshifts
are sharply peaked around $z = 10$. The photometric probability distributions
 are included in Figure \ref{fig1} alongside
the images of the candidates.

\begin{figure}
\centering
\epsscale{1.2}
\plotone{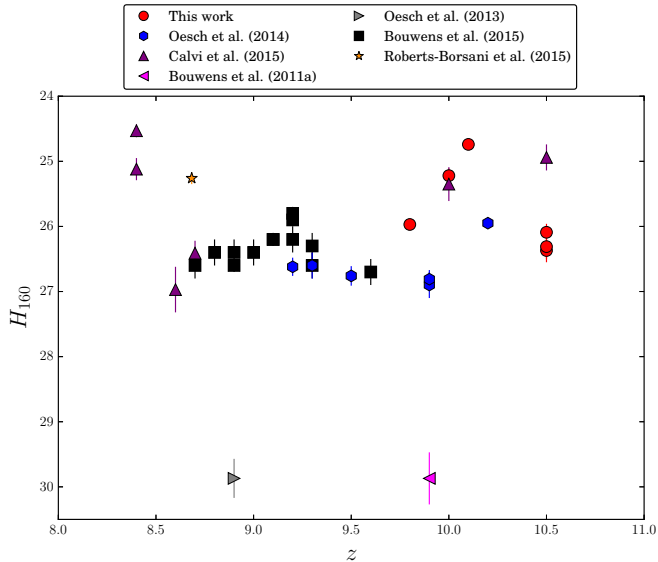}
\caption{The apparent \hband-band magnitude vs. redshift for $z \ge 8$ galaxy candidates. The red circle points refer to the candidates described in Sections \ref{cand129}-\ref{cand785}. Other points refer to candidates from other \emph{HST}/WFC3 surveys as labelled. The redshift $z$ is the photometric redshift for all candidates except that from \citet{robertsborsani15}, where we use the spectroscopic redshift from \citet{zitrin15}. \label{hvszplot}}
\end{figure}

A comparison of the apparent \hband{} 
magnitude against the photometric redshift of our candidates against $z \ge 8$ candidates 
from other \emph{HST}/WFC3 surveys is shown in Figure \ref{hvszplot}. While two of our candidates are particularly bright in \hband, 
they are consistent with previously-discovered candidates at $z \sim 10$ by \citet{calvi15}.

We also determine the size of the candidates, starting from the
observed half-light (effective) radius as determined by
\texttt{SExtractor}, which is translated into an intrinsic source size
taking into account the effects of the point-spread function (PSF) broadening and surface
brightness limits following \citet{calvi15}. The empirical
  relation has been constructed by inserting and recovering artificial
  sources with known input size and magnitude into BoRG
  images. Source size is very helpful to help discriminate between high- and low-$z$  sources, since direct measurements by \citet{holwerda14} on CANDELS galaxies show that $z> 9- 10$ sources are more compact than $z \sim2$ contaminants
with similar colors. This empirical separation might be related to an approximate scaling of galaxy sizes as $(1+z)^n$ with $n\sim-1$ \citep{fall80,bouwens04,bouwens06,oesch10}, although a recent study by \citet{curtislake16} highlights that the intrinsic sizes likely evolve less strongly with redshift ($n\sim-0.2$) compared to observed sizes.
We discuss the contamination of our
sample further in Section \ref{contamination}.

\subsection{borg\_0240-1857\_129}\label{cand129}
This candidate is the brightest in the sample, with magnitude
$H_{160} = 24.7$. It is robustly detected in \hband{} at
$\mathrm{S/N} = 14.5$, and marginally detected in \jband{} at
$\mathrm{S/N} = 2.5$, even though it is close to the edge of the
chip. The source has a very red $J_{125} - H_{160}$ color, with
$J_{125} - H_{160} = 2.2$. It also shows extended structure, and has
$r_e = 0\farcs33$. Its photometric redshift solution is
sharply peaked at $z = 10.1$, with a broad higher-redshift wing.

\subsection{borg\_0240-1857\_369}
This candidate, in the same field as the previous one is detected with
magnitude $H_{160} = 25.2$, making it the second-brightest source in
the sample. It is detected with S/N $= 9.6$ in \hband, and again
marginally detected with S/N $ = 2.2$ in \jband.  It is the most
extended source in the sample, with $r_e = 0\farcs38$.  Its
photometric redshift, like borg\_0240-1857\_129, is peaked at $z = 10.0$,
with a broad higher-redshift wing.

\subsection{borg\_0240-1857\_25}
Field borg\_0240-1857 includes a third bright candidate with $H_{160} = 26.4$, detected at
S/N $=8.1$. This source is not detected in the other
bands (\jband, \yband{} or \vband$_{600}$). Unlike the two brighter
candidates, this object is more compact, with a measured half-light
radius $r_e = 0\farcs13$. This is smaller than the PSF of
the image ($0\farcs15$), indicating that it could be a
point-source-like contaminant such as a cool dwarf star, although the
stellarity of this source is $0.71$, which is lower than the value
expected for a point source  (e.g. \citealt{schmidt14} uses
CLASS\_STAR $< 0.85$ and \citealt{bouwens15a} CLASS\_STAR $<0.9$ to
exclude stars).
 
This candidate is close to a foreground galaxy with $H_{160} = 26.0$,
with a centre-to-centre projected separation of $1\farcs25$. While
this foreground galaxy has an uncertain photometric redshift solution,
it is likely to be at $z > 0.5$, based on its compact size. Using the
framework developed by \citet{baronenugent15a} and \citet{mason15}, we
estimate the gravitational lensing of this source.  Magnification PDFs are obtained by estimating velocity dispersions from $H_{160}$ magnitudes, using the empirical redshift-dependent Faber-Jackson relations given in \citet{mason15} and \citet{baronenugent15a}. Velocity dispersion is the best tracer of the strength of a strong gravitational lens \citep{turner84,schneider06,treu10}. The Einstein radii of the foreground objects are modelled as singular isothermal spheres (e.g. \citealt{treu10}) which depend on the velocity dispersion and the angular diameter distance to the source, and between the lens and source (where we use the best photo-$z$ values). Assuming that the
foreground source is at $z \sim 2$ (which maximizes lensing
magnification), we infer a magnification $\mu = 1.1 \pm 0.1$.

\subsection{borg\_0456-2203\_1091}

This object has a magnitude $H_{160} = 26.1$ ($\mathrm{S/N} = 8.1$),
and is detected in the \hband{} only, with an extended but compact
structure (effective radius $r_e = 0\farcs24$).

The source is located relatively close ($0\farcs5$
separation) to a hot pixel, which appears in the \yband{} and \jband{}
images. The \hband-band image is unaffected since it was acquired in a
later orbit than the images in bluer bands. We carefully examined the
individual FLT files and conclude that since the separation between
the source center and the hot pixel is larger
than twice $r_e$, and there is no sign of a hot pixel in the \hband-band,
the identification of the candidate as a \jband-dropout is robust.

\subsection{borg\_1153+0056\_514}

This candidate is detected with a magnitude $H_{160} = 26.3$, and has
S/N $=8.0$. It is not detected in \jband, \yband{} or \vband. It has
an effective radius of $r_e = 0\farcs23$. This candidate is close to 
a foreground object ($1\farcs46$
centre-to-centre projected separation). The foreground object has an
apparent magnitude $H_{160} = 25.0$, and is at an indeterminate
photometric redshift
. We use the same modelling framework as for borg\_0240-1857\_25 to estimate the lensing magnification of this source. Assuming that the source is at $z \sim 2$, we find a maximum $\mu = 1.2 \pm 0.1$.

Analysis of the FLT images of this field highlighted the presence of a
bad pixel, correctly identified and masked by the data reduction
pipeline, at the outer edge of the segmentation map of the dropout
candidate in one of the two \hband{} frames. To determine the impact
on the final photometry, we measured the source flux in the FLT frames
separately, finding that the candidate is detected with S/N
$ = 5.1$ in the unaffected image and also S/N $= 5.1$ in the image
affected by the bad pixel. Hence, we are confident that the source is
real and that the photometry from the final drizzled image is robust.

\subsection{borg\_1459+7146\_785}\label{cand785}

The sixth and final candidate is confidently detected at S/N$=12.8$ in \hband{}
($H_{160} = 26.0$), and also in the \jband{} with S/N = 3.7. Its
photometric redshift is sharply peaked at $z = 9.8$, with a secondary
solution at $z \sim 2.5$.  This candidate is also very compact, with
measured half-light radius $r_e = 0\farcs14$, and the
highest stellarity of the sample (CLASS\_STAR = 0.91). Combining
compactness with high stellarity from a high S/N source, a stellar
nature (cool dwarf) for this source is relatively likely, as we
discuss in Section~\ref{contamination}.

\section{Number density and luminosity function of $\lowercase{z}\sim 10$
  galaxies}\label{lfsection}

To translate the results on the search of possible candidates at
$z\sim 10$ from the archival BoRG[z8] data into a number
density/luminosity function determination, we need to assess both the
impact of contamination in our sample, and the effective volume probed
by the data.

\subsection{Contamination}\label{contamination}

There are multiple classes of lower-$z$ sources that may have similar
$J_{125} - H_{160}$ colors to $z \sim 9-10$ Lyman-break galaxies (LBGs), such as Galactic
stars, intermediate-redshift passive galaxies, and strong line
emitters.

Cool, red stars in the Milky Way may be possible contaminants of our
sample, although typical colors lack a strong $J_{125} - H_{160}$
drop. At low signal-to-noise ratio, the separation of point-like Galactic
stars from resolved galaxies using the \texttt{SExtractor} CLASS\_STAR
parameter is not fully reliable. We use CLASS\_STAR $< 0.95$ in our
selection of \jband-dropouts in Section \ref{section3} to reject
artifacts remaining from the reduction process, but this is not a
strict enough criterion to reject all stars from our sample. In this
case, five of our candidates identified as
\jband-dropouts have CLASS\_STAR $< 0.8$, with only borg\_1459+7146\_785
having CLASS\_STAR $> 0.9$ (a value considered by \citealt{bouwens15a}
as indicative of a stellar nature). Therefore we conclude that this
source is most likely a stellar contaminant with unusual
colors. 

Emission-line galaxies are another source of contamination for
$z \sim 9-10$ galaxy samples. For example, galaxies at $z \sim 3$ with
strong [OII] emission may appear bright in \hband-band while the
galaxy continuum is too faint to be detected in the other bands
\citep{atek11, huang15}. \citet{bouwens15b} find that, at $z \sim 8$,
the average density of extreme line emitters that enter the photometric selection is $\sim 10^{-3}$ per
arcmin$^{-2}$, by creating mock catalogs of extreme emission line galaxies with varying $J_{125}$ magnitude and spectral slope $\beta$. Extrapolating this result to $z \sim 10$, we expect to
find $n_{\mathrm{c}} = 0.3$ potential contaminants of this type over our survey
area. This value is in line with previous spectroscopic observations
of $z \sim 8$ BoRG candidates by \citet{treu12,treu13} using the
MOSFIRE spectrograph on the Keck telescope, and by \citet{baronenugent15} using XSHOOTER on the Very Large Telescope. These studies found no emission lines in the spectroscopy of $z \sim 8$ candidates, and are able to rule out emission lines from a low-$z$ extreme emission line contaminant to $5\sigma$, assuming that all of the $H_{160}$ flux is due to a strong emission line. \citet{baronenugent15} also find that, with a 3hr exposure, only a small part of the spectrum ($\sim14\%$) is so affected by atmospheric transmission and absorption by OH lines that a strong emission line would not be detected to $2\sigma$. 

The last, and probably most severe class of contaminants, is that of
passive and dusty galaxies that thus show a strong Balmer break and a
very faint UV continuum. Under these conditions, $z\sim 2$ sources may
mimic properties of LBGs and thus enter into our
selection. Observations with \emph{Spitzer}/IRAC at $3.6$ and
$4.5\,\mu$m can efficiently distinguish between high- and low-redshift
sources. In fact, dusty $z \sim 2$ galaxies will appear 1-2 magnitudes
brighter in [3.6] and [4.5] than in \hband, while $z \sim 9-10$ LBGs
will have a relatively flat spectrum. Without Spitzer data, we rely on
the size of the sources as proxy for the $H_{160}-[4.5]$ color,
considering sources with $r_e>0\farcs3$ as likely contaminants. 
\citet{holwerda14} find that, while the mean size of candidates in the $z \sim 10$ sample from \citet{oesch14} is $0\farcs13$, low-redshift, IRAC-red interlopers have a mean size of $0\farcs6$, but can be as small as $0\farcs35$, and there are no high-$z$ candidates with sizes greater than $0\farcs2$ (Figure 4, \citealt{holwerda14}). Hence, we take $0\farcs3$ as a threshold. The two brightest sources in our sample are so
extended to fall into such classification. The sources considered in \citet{holwerda14} are fainter than the $z \sim 10$ candidates in our sample, and so it is conceivable that the larger sizes of borg\_0240-1857\_129 and borg\_0240-1857\_369 are due to their higher luminosities. Using the size-luminosity relation derived in \citet{holwerda14}, we infer that the size of a $z=10$ galaxy at $M_{\mathrm{AB}}=-22.7$ (the brightest in our sample) would be $0\farcs17 \pm0\farcs04$, below our threshold of $0\farcs3$. This size cannot be used to completely reject extreme emission-line galaxies, which are likely to be more compact. For example, \citet{huang15} find that their sample of 52 extreme emission line galaxies in CLASH have FWHM $< 0\farcs9$, similar to our $r_e < 0\farcs3$ criterion for $z \sim 10$ candidates. 

In addition to finding the redshift probability distribution of our six candidates using 
BPZ, we also fit SED templates described in \citet{oesch07}. From this, we find an average probability 
$P(z < 8)$ of $54\%$. We conclude that three out of the six
candidates, borg\_1459+7146\_785, borg\_0240-1857\_129, and borg\_0240-1857\_369, are likely to be contaminants. For the remaining
three, contamination is still a possibility, and hence we make a
conservative assumption of $33\%$ contamination  (two out of three sources at $z\ge10$),
based on the estimate from the BoRG[z9-10] survey \citep{calvi15} using the average probability  $P(z < 8)$ for the candidates in their sample.

\subsection{Clustering in borg\_0240-1857: evidence for or against contamination? }

Of the six possible candidates identified in the full BoRG[z8] survey,
three of them, including the two brightest, are found in the same
field (borg\_0240-1857). The exposure time for this pointing is
similar to the median of the survey ($t_{\mathrm{exp}}=1400$ s in
\hband{}), and thus such a catch is highly unlikely under the
assumptions of a uniform distribution of candidates in the 62 fields
analyzed here. Based on theoretical expectations, the presence of
clustering can be used to verify the identity of bright high-$z$
sources \citep{munoz_loeb08}, under the broad assumption that UV
luminosity is correlated with dark-matter halo mass (e.g.,
\citealt{trenti10,tacchella13}). Overdensities have also been
identified at $z\sim 8$ in LBG samples
\citep{trenti12,ishigaki15b}. However, one alternative possibility,
more consistent with the relatively large size of the two bright
$J$-dropouts, would be the presence of an overdensity of passive and
dusty satellite galaxies within an intermediate redshift group. In
either case, a further exploration of this configuration is very
interesting since it can either identify an exciting overdensity of
unexpectedly bright sources at $z\sim10$, or shed light on the
properties of intermediate redshift galaxies with extreme $J-H$
colors.

\subsection{Completeness}\label{completeness}

\begin{figure}
\centering
\epsscale{1.1}
\plotone{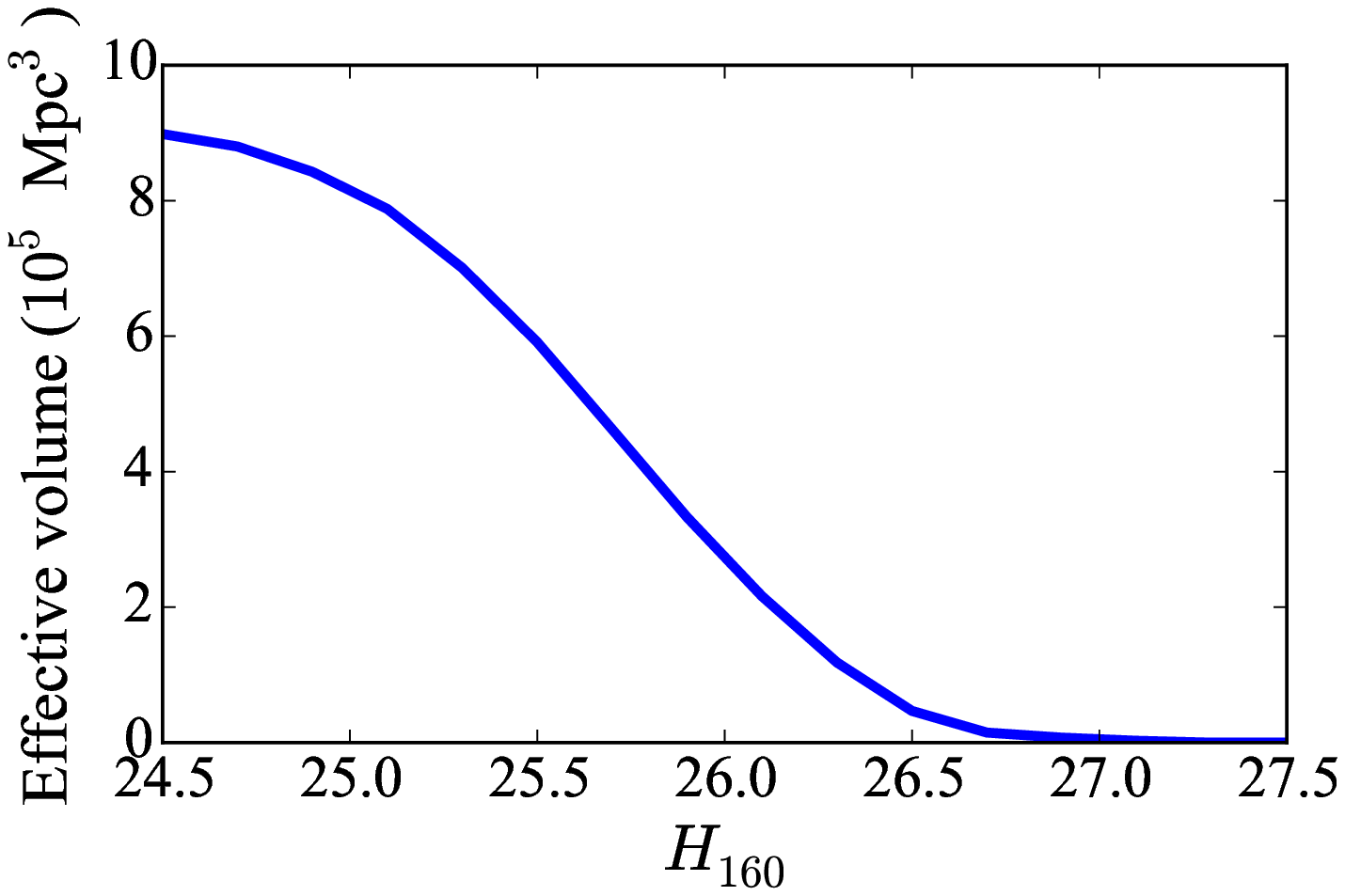}\\
\includegraphics[width=0.46\textwidth,natwidth=610,natheight=642]{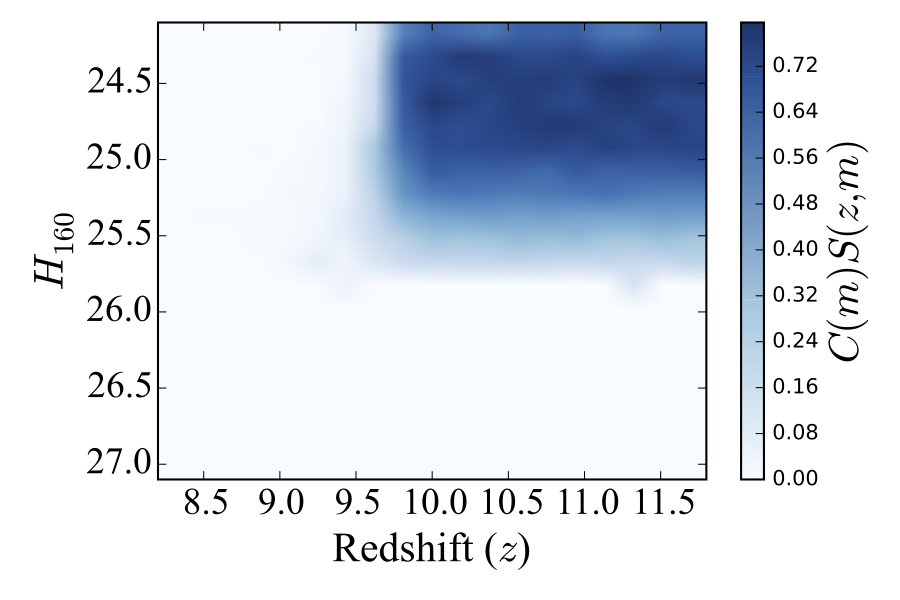}
\caption{Top panel: The effective volume in comoving Mpc$^3$ recovered from simulations, as a function of the apparent \hband{} magnitude. Bottom panel:  An example of the selection function $S(z,m)$ for field borg\_0440-5244. The selection function is derived from simulations, by inserting and recovering artificial sources. \label{figvolume}}
\end{figure}

We perform source recovery simulations to determine the efficiency and
completeness of our selection, following \citet{oesch07,oesch09, oesch12}. 
To do this, we insert and recover artificial
galaxies with a S\'ersic luminosity profile in the
images. Half of the sources follow a de Vaucouleurs profile (S\'ersic
index $n=4$), and the other half follow an exponential profile
(S\'ersic index $n=1$), spanning a range of magnitudes
($24 \le H_{160} \le 28$), redshifts ($8.2\le z \le 11.8$) and
sizes (logarithmic distribution with mean 0\farcs175 at $z\sim7$, scaling
as $(1+z)^{-1}$). The spectra of the sources are modeled as power law
$f(\lambda)\propto \lambda^{\beta}$ with $\beta = -2.2 \pm 0.4$ (Gaussian distribution) 
with a sharp cut-off at rest-frame $\lambda =1216~\mathrm{\AA}$. 
The intrinsic profiles of the artificial sources
are convolved with the WFC3 PSF for each corresponding filter, before
being inserted into the BoRG science images at random
positions. Sources are then identified with {\tt{SExtractor}}, and the
statistics of the recovery rate is quantified. This is through the
definition of $C(m)$ which is the completeness of the source
detection, that is the probability of recovering an artificial source
of magnitude $m$ in the image, and of $S(z,m)$, which is the
probability of identifying an artificial source of magnitude $m$ and
redshift $z$ within the dropout sample, assuming that the source
  is detected. One example of the selection function for the dropout
search in field borg\_0440-5244 is shown in the bottom panel of
Figure \ref{figvolume}, while the upper panel of the same figure shows
the overall effective volume probed by our search over all BoRG
archival fields as a function of the apparent \hband{} magnitude.

\subsection{Determination of the luminosity function}\label{lfdetsection}

\begin{figure}
\epsscale{1.2}
\plotone{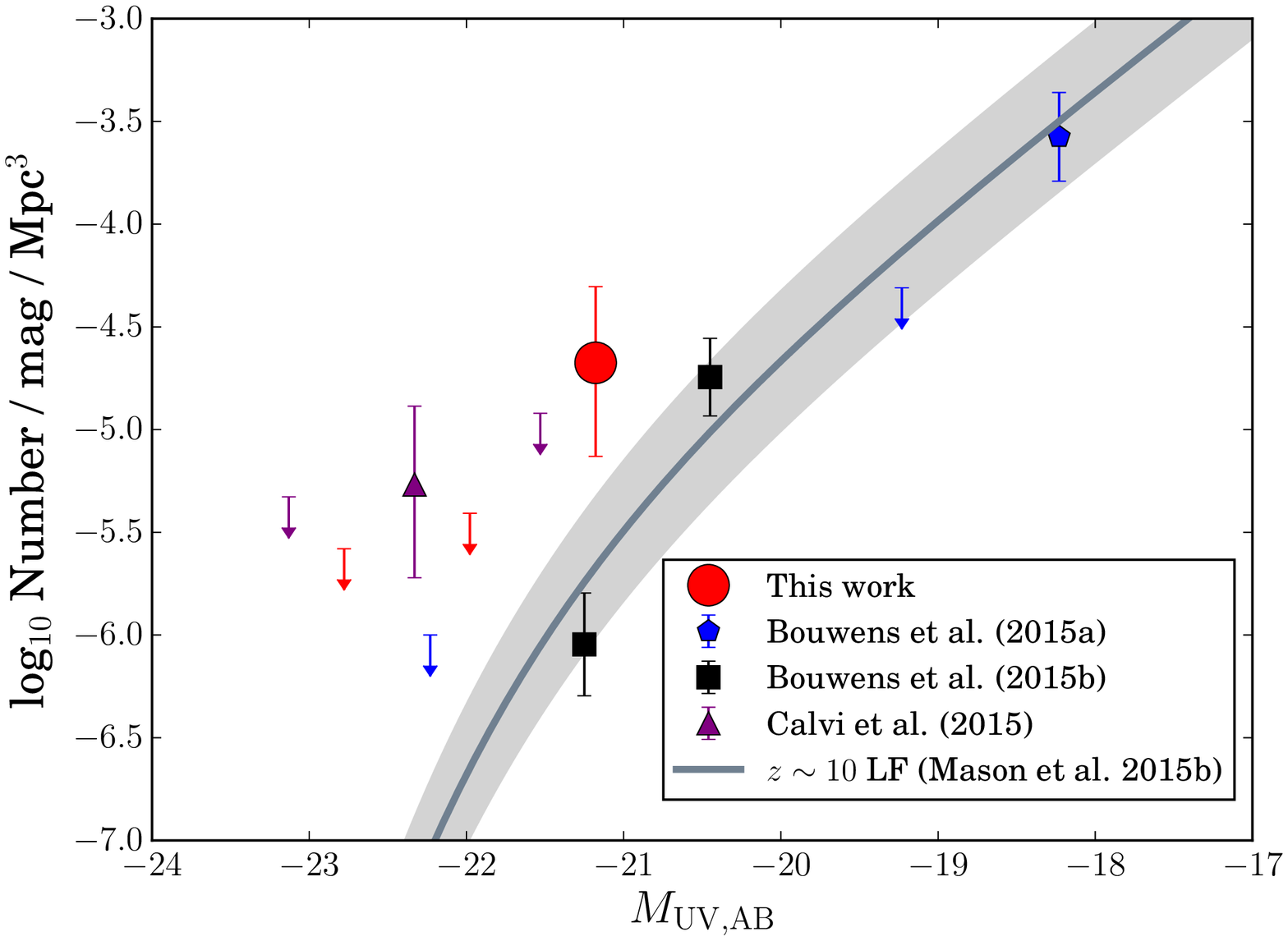}
\caption{Step-wise determination of the UV LF at $z \sim 10$. The red
  circle and red upper limits refer to the values discussed in Section
  \ref{lfdetsection}. Other symbols refer to
  \citet{bouwens15a,bouwens15b,calvi15} as labelled. Error bars are
  $1\sigma$ Poisson errors, and limits are $1\sigma$ upper limits. Our
  best fit determination is shown as solid red line, while the dashed
  red line is one example of another equally acceptable fit,
  highlighting that the current data are insufficient for strong
  constraints on the LF shape. The overplotted gray line indicates the
  $z \sim 10$ LF from the theoretical model of \citet{mason15b}, with
  shaded region being the 68\% contour of its $\phi_*$
  uncertainty. \label{figLF}}
\end{figure}

From the discussion in Section~\ref{contamination}, we consider the
two brightest sources to be likely contaminants because of their large
half-light radii, and we exclude the point-like source
borg\_1459+7146\_785 as well. For the surviving three candidates we
assume a contamination rate of $33\%$, e.g. we expect two sources to
be at $z\sim 10$. After taking into account the effective volume
probed by our selection, our estimates for the bright-end of the
luminosity function at $z\sim 10$ is reported in Table~\ref{table3},
and shown in Figure~\ref{figLF}. Interestingly, we infer a higher
number density of bright sources than previous determinations by
\citet{bouwens15a,bouwens15b} around $M_{AB}\sim -21$, although the
uncertainty is very large because of the small number of
candidates. For brighter sources ($M_{AB}\lesssim -21.5$), our upper
limits on $z\sim 10$ density are similar to those obtained in legacy
fields, and strengthen the evidence for suppression of the abundance
of galaxies at the bright-end of the luminosity function.

When
compared to the initial results from the ongoing BoRG[z9-10] survey
\citep{calvi15}, assuming that our two brightest sources are low-redshift contaminants, we do not find evidence of ultra-bright
($M_{AB} \sim -22$) galaxies despite analyzing data covering more than
twice the area. If follow-up observations of our brightest sources indicate that they are likely at high redshift, we would instead determine that the LF is higher at the bright end than the upper limit from \citet{bouwens15a}, and is instead consistent with the determination by \citet{calvi15} at $M_{AB} = -22.3$.

Overall, our LF determination is higher, but still consistent at
$\sim 2\sigma$ with the theoretical model of \citet{mason15b}, shown
as grey shaded area in Figure~\ref{figLF}. Previous studies did not
attempt unconstrained fits to the $z\sim 10$ LF, likely because of the
small number of candidates. To evaluate the status of the situation
with our additional datapoints, we attempt to derive Schechter parameters for a
maximum likelihood fit to the stepwise LF data, carried out assuming a
Poisson distribution for the number of galaxies expected in each bin
(see \citealt{bradley12}). Due to the non-detection at $M_{\mathrm{AB}} = -19.23$ by \citet{bouwens15a}, the LF 
is suppressed at the faint end. This leads to a likelihood landscape that is
very flat over a wide region of the parameter space, and hence, we are unable to sufficiently constrain the Schechter parameters. 
Our fit attempt thus highlights that the dataset is
still too small for tight quantitative constraints, but future growth
in the number of candidates identified will allow rapid improvements.

Finally, we note that our conclusions rest on the assumption that the
two brightest candidates we identified in field borg\_0240-1857 are
contaminants. If we were to include them in the analysis as $z\sim 10$
sources, we would infer that the LF would favor a power law at the
bright end, rather than a Schechter form. Evidence for a single or double power-law
form at high redshift has been seen in determinations of the LF at
$z\sim7-8$ (\citealt{bowler14,finkelstein15}, also earlier considered by \citealt{bouwens11b}), and potentially at
$z\sim10$ by \citet{calvi15}, and may be naturally interpreted as a
decrease in mass quenching from processes such as AGN feedback at high
redshift \citep{bowler14}. Magnification bias, however, can also produce this effect on an otherwise Schechter-like LF. 
Thus, the astrophysical interpretation of
our search ultimately rests on follow-up observations to establish the
nature of the candidates borg\_0240-1857\_129 and
borg\_0240-1857\_369. In any case, it is very interesting to note that
the number of potential candidate $J$-dropouts that we identified is
small (just six in over 60 fields), making further observations
time-efficient, especially because half of the sources are clustered
in a single pointing.

\begin{deluxetable}{cc}
\centering
\tabletypesize{\footnotesize}
\tablecolumns{12}
\tablewidth{0pt}
\tablecaption{Step-wise rest-frame UV LF at $z \sim 10$ \label{table3}}
\tablehead{ \colhead{$M_{UV,AB}$}  & \colhead{$\phi \,(10^{-5} \,\mathrm{Mpc}^{-3} \,\mathrm{mag}^{-1})$}}
\startdata
$-22.78$ & $< 0.26$ \\
$-21.98$ & $< 0.39$\\
$-21.18$ & $2.1^{+2.9}_{-1.4}$
\enddata
\end{deluxetable}

\section{Conclusions}\label{sec:conclusions}

In this paper we presented a search for $z \sim 9-10$ candidates in
archival data of the 2010-2014 Brightest of Reionizing Galaxies
(BoRG[z8]) survey, a pure-parallel optical and near-infrared survey
using \emph{HST}/WFC3. While the survey was optimized to identify
$z\sim 8$ sources as \yband-dropouts, we searched over the deepest 293
arcmin$^2$ of the survey for \jband-dropout sources with $H_{160} \lesssim 26$,
motivated by recent identification of very bright sources with
$z\sim 9$. Our key results are:

\begin{itemize}

\item We identify six $z \sim 10$ galaxy candidates, detected in \hband{} at S/N $>8$ and
  satisfying a conservative $J_{125}-H_{160}$ color selection with non-detection in bluer
  bands of the survey. The candidate's magnitudes vary from $H_{160}=24.7$ to
  $H_{160}=26.4$. Analysis of the surface brightness profile leads to the
  tentative identification of three contaminants, with the two
  brightest sources likely being intermediate redshift passive
  galaxies due to their size, and one faint source a galactic cool
  dwarf star because of the compact size and high stellarity.

\item Of the six candidates, three are in the same field, borg\_0240-1857,
  including the two brightest of the sample. Such strong clustering
  would be naturally explained if the sources were $z\sim 10$ (see
  \citealt{munoz_loeb08}), despite contrary indication from $r_e$, but an
  alternative explanation of sub-halo clustering at intermediate
  redshift would also be viable.

\item Based on our best estimate of the LF, we infer a higher galaxy number
density for sources at $M_{AB}\sim -21$ compared to the observations
of Bouwens et al. (2015a,b) and with the theoretical model of Mason et
al. (2015b). However, our measurement is still consistent at the
$2\sigma$ level with these studies.

\item Irrespective of the nature of the two brightest sources in the
  sample, the selection criteria that we adopted yield a small number
  of candidates, very manageable for follow-up observations. This is
  quite remarkable, since the BoRG[z8] survey was not designed with
  $z\sim10$ in mind, and the number of contaminants could have been
  much larger given the absence of a second detection band and the
  lack of a near-UV color to help remove passive and dusty
  intermediate redshift galaxies. 

\item Targeted follow-up observations can efficiently clarify the
  nature of the candidates we identified, help to further constrain
  the bright-end of the LF and characterize the properties of the yet
  unstudied population of compact intermediate redshift passive
  galaxies that mimic the colors of $z>8$ sources.
\end{itemize}

The efficiency of targeted follow-ups and the overall potential to
complement searches for $z\sim 10$ sources traditionally
carried out in legacy fields are demonstrated by the very recent award
of \emph{Spitzer} IRAC time to our team to investigate the nature of the
sources discussed here (PID \#12058, PI Bouwens). With these observations, it will be possible to
clarify the behavior of the bright end of the LF at
$z\sim 10$, as well as to confirm ideal targets for further
spectroscopic follow-up investigations without having to wait for
\emph{James Webb Space Telescope}.

\acknowledgements

SB, MT, and JW thank the Vatican Observatory, where part of this work
was carried out, for kind hospitality. This work was partially
supported by grants {\it HST}/GO 13767, 12905, and 12572. This research was conducted by the Australian Research Council Centre of Excellence for All-sky Astrophysics (CAASTRO), through project number CE110001020.

\end{document}